\documentclass[12pt,letterpaper]{article}
\usepackage[dvipsnames]{xcolor}
\usepackage[margin=1in,footskip=0.5in]{geometry}
\usepackage[T1]{fontenc}
\usepackage{verbatim}
\usepackage{amsthm}
\usepackage{amsmath}
\usepackage{booktabs}
\usepackage{amssymb}
\usepackage{caption}
\usepackage{natbib}
\usepackage{lmodern}
\usepackage{soul}
\usepackage{bbm}
\usepackage{newcent} 
\usepackage{setspace}
\usepackage{subfigure}
\usepackage{xcolor}
\usepackage{graphics}
\usepackage{epsfig}
\usepackage{geometry}
\usepackage{accents}

\usepackage{enumitem}
\usepackage{setspace}
\usepackage{array}
\newcolumntype{x}[1]{>{\centering\arraybackslash\hspace{0pt}}p{#1}}
\usepackage{tikz}
\usetikzlibrary{patterns}
\usetikzlibrary{patterns.meta}
\pgfdeclarepattern{
  name=hatch,
  parameters={\hatchsize,\hatchangle,\hatchlinewidth},
  bottom left={\pgfpoint{-.1pt}{-.1pt}},
  top right={\pgfpoint{\hatchsize+.1pt}{\hatchsize+.1pt}},
  tile size={\pgfpoint{\hatchsize}{\hatchsize}},
  tile transformation={\pgftransformrotate{\hatchangle}},
  code={
    \pgfsetlinewidth{\hatchlinewidth}
    \pgfpathmoveto{\pgfpoint{-.1pt}{-.1pt}}
    \pgfpathlineto{\pgfpoint{\hatchsize+.1pt}{\hatchsize+.1pt}}
    \pgfpathmoveto{\pgfpoint{-.1pt}{\hatchsize+.1pt}}
    \pgfpathlineto{\pgfpoint{\hatchsize+.1pt}{-.1pt}}
    \pgfusepath{stroke}
  }
}

\tikzset{
  hatch size/.store in=\hatchsize,
  hatch angle/.store in=\hatchangle,
  hatch line width/.store in=\hatchlinewidth,
  hatch size=5pt,
  hatch angle=0pt,
  hatch line width=.5pt,
}
\usepackage{pifont}

\usepackage{avant}

\newcolumntype{L}[1]{>{\raggedright\let\newline\\\arraybackslash\hspace{0pt}}m{#1}}
\newcolumntype{C}[1]{>{\centering\let\newline\\\arraybackslash\hspace{0pt}}m{#1}}
\newcolumntype{R}[1]{>{\raggedleft\let\newline\\\arraybackslash\hspace{0pt}}m{#1}}

\makeatletter \renewenvironment{proof}[1][\proofname]
{\par\pushQED{\qed}\normalfont\topsep6\p@\@plus6\p@\relax\trivlist\item[\hskip\labelsep\bfseries#1\@addpunct{.}]\ignorespaces}{\popQED\endtrivlist\@endpefalse} \makeatother

\theoremstyle{plain}
\newtheorem{theorem}{Theorem}
\newtheorem{lemma}{Lemma}

\newtheorem{proposition}{Proposition}

\usepackage{multirow}
\usepackage{chngcntr}
\usepackage{apptools}
\AtAppendix{\counterwithin{lemma}{section}}
\AtAppendix{\counterwithin{proposition}{section}}

\definecolor{uncblue}{RGB}{75, 165, 211}
\definecolor{nberblue}{RGB}{0, 90, 155}
\definecolor{resgreen}{RGB}{34, 84, 67}
\definecolor{pennblue}{RGB}{0, 44, 119}
\definecolor{pennred}{RGB}{152, 30, 50}

\usepackage{hyperref}
\hypersetup{%backref,
hypertexnames=false,
pdfborder = {0 0 0},
urlbordercolor = {0 0 0},
colorlinks=true,
linkcolor=pennblue,
urlcolor=nberblue,%pennred,
citecolor=pennblue}
\begin{document}
\newcommand{\state}{\ensuremath{\theta}}

\newcommand{\cert}{\ensuremath{e}}

\newcommand{\type}{\ensuremath{\mu}}

\newcommand{\E}{\ensuremath{\mathbb{E}}}

\newcommand{\R}{\ensuremath{\mathbb{R}}}

\title{Brave New Data: Pricing Expertise and Discovery}
\author{Dihan Zou\thanks{Department of Economics, University of North Carolina at Chapel Hill. Email: \href{mailto:dihan@email.unc.edu}{\texttt{dihan@email.unc.edu}}. I thank 
Gary Biglaiser,
Fei Li,
Genyu Li,
Peter Norman,
Michael P.H.\ Tam,
and Jiaming Wei
for helpful comments. All errors are my own.}}
\date{\today}

\maketitle
\begin{abstract}
We study the procurement of verifiable evidence from an expert
who privately assesses a binary state and chooses the direction
and intensity of evidence acquisition.
The buyer balances the ease of inducing discovery against its value for decisions beyond the expert's initial assessment.
An optimal contract pays only for disclosed evidence,
independently of the initial report. 
When acquisition is cheap,
equal rewards for evidence of either state induce intensive
search whose failure is informative. 
As costs rise, rewarding evidence of only one state becomes optimal despite the symmetry of the environment, 
reducing expected payments at the expense
of decision accuracy.
\end{abstract}

\medskip

{\small 

}

\newpage

\tableofcontents

\newpage

\section{Introduction}
Data buyers commission experts to produce and validate specialized data.
For example, \mbox{OpenAI} engaged professional software developers
to assess and review software engineering tasks for inclusion in SWE-bench
Verified, an expert-validated benchmark for evaluating the coding
capabilities of large language models.
Other examples include Epoch AI's contracts with mathematicians
to create original problems for FrontierMath and Premise's payments
to local contributors to collect geospatial data.\footnote{
\citet{OpenAI2024SWEbenchVerified} describes its initial
concerns about benchmark reliability and the subsequent review
by professional software developers.
\citet{EpochAI2026FrontierMath} documents OpenAI's commission
and the contracts with researchers
to produce problems for the Tier~4 expansion.
For Premise's paid task marketplace and an application involving
field visits to locate health facilities in Sierra Leone, see
\citet{PremiseGeospatialCaseStudy}.
}
Such work can build on knowledge that contributors already possess.
Familiarity with a question can suggest a plausible answer and
guide further reasoning, computation, or investigation toward
promising ways to establish it. Additional work can develop this
preliminary understanding into verifiable evidence or reveal
grounds for revising it. Existing knowledge and further investigation
thus jointly contribute to the production of reliable data.
How should a data buyer compensate an expert for this combination
of expertise and discovery?

This paper studies a data procurement setting in which an expert provides an initial assessment and undertakes further investigation.
Consider a firm deciding whether to include a task in an AI evaluation dataset. 
An expert privately assesses whether the task
reliably measures the intended ability and reports this assessment
before investigating further. 
Her report can already inform the firm's inclusion decision, while additional work can produce verifiable evidence about the task's suitability.
Existing knowledge therefore plays two roles: it guides the expert toward findings she expects to establish and determines how much additional evidence can improve the buyer's decision. 
These considerations need not favor the same investigation. 
The expert
weighs expected rewards against investigation costs, whereas the
buyer values the improvement in his decision relative to what her initial assessment already provides.
Tailoring rewards to the reported assessment introduces a further
difficulty: 
because the underlying assessment is private, the
compensation scheme also affects how the expert reports what she
knows. 
The contracting problem is therefore to use existing expertise to guide valuable discovery while respecting the incentives to communicate that expertise and undertake further investigation.

We study this problem in a model with two states and two actions. A data buyer (the principal) wishes to match his action to the state. 
An expert (the agent) has a private initial assessment of the
probability of one state. This informed prior is her type.
The principal commits to a deterministic contract specifying nonnegative payments based on her reported assessment and any evidence she subsequently discloses. 
The agent then reports her assessment and privately chooses which state to investigate and how much effort to exert. 
Investigation can produce a certificate: verifiable evidence establishing that the targeted state is true. 
Greater effort increases the probability of finding this certificate when that state is true. 
Acquired evidence can be withheld but cannot be fabricated. 
The principal observes the initial report and either one certificate or silence (no disclosure), and then chooses his action optimally. 
On the other hand,
the agent values compensation net of effort cost and has no intrinsic preference over the decision. 
The conflict over investigation therefore arises from the interaction between
reporting private expertise and earning rewards for discovery.

We first consider a benchmark in which the principal observes the agent's type, while her search target and effort may remain hidden. 
In this benchmark, it is optimal to investigate the state the agent initially considers less likely.
A certificate of that state overturns the
action favored by existing knowledge, while unsuccessful search
reinforces it. When the principal controls acquisition, he reimburses
its cost. 
When search is hidden, he must instead reward discovery, leaving the agent an expected payment above her effort cost. 
This raises the cost of inducing effort and weakly reduces its optimal level, but preserves the preferred direction of search.

Private expertise creates an additional distortion.
Payment design must account for an agent's incentive to imitate
another agent's report and subsequent search choices.
For any given payment terms, an increase in the probability of
one state makes searching for its certificate more attractive
and searching for the other state's certificate less attractive.
Joint incentive compatibility therefore requires that those investigating a given state
consider it at least as likely as do those investigating
the other state.
As a result, the principal cannot induce every agent to investigate her less likely state, although he can induce some agents to do so.
The same conflict shapes discovery bonuses: the additional payment for a certificate relative to the payment when no certificate is disclosed. Consider two distinct types with the same search target.
To maintain truthful reporting, the type assigning a higher
probability to the targeted state must receive a weakly higher
discovery bonus.
The principal's relative preference for a larger bonus moves in the opposite direction: stronger incentives are more attractive
when the targeted state is less likely.
These opposing orderings favor offering the same discovery bonus to all agents searching for the same certificate. 

We show that an optimal contract can be chosen to pay nothing when no certificate is disclosed and to offer either equal rewards for both certificates or a reward for only one.
These payments do not depend on the initial report, although the principal uses that report when interpreting silence and choosing his action.
The choice between these payment structures balances the ease of
inducing discovery against its value for the principal's decision.
At a given bonus, an agent searches more intensively when she
considers the rewarded state more likely, until effort reaches its
upper bound. The greater chance of success also raises the frequency
of payment. When that state is already favored by the initial
assessment, however, its certificate confirms the action the principal
would otherwise take. Such search improves the decision
only if effort makes silence sufficiently informative to overturn
that assessment. Evidence of the less likely state instead changes
the decision whenever it arrives, but its lower probability weakens
the incentive to search. Stronger rewards can induce more informative
investigation, while increasing payments on all rewarded discoveries,
including those that confirm existing knowledge. The principal
therefore chooses which certificates to reward and the size of their
bonuses to balance the reduction in decision loss against expected
payments.

When acquisition is sufficiently cheap, equal rewards for both
certificates are optimal. Each agent searches the state she considers
more likely, which offers the greater chance of earning the reward.
Cheap acquisition makes this direction of search compatible with
informative investigation: a modest bonus induces sufficiently
intensive search that failure to find the expected certificate can
overturn the initial assessment. At full effort, either a certificate
or its absence reveals the state. Accommodating the agent's preferred
target therefore need not sacrifice much decision accuracy when
search is intensive. The value of investigation comes from both its
discoveries and what the principal learns when it fails.

As acquisition becomes more costly, the optimal payment structure
switches to rewarding only one certificate, even though the two
states are symmetric in the primitives. This redirects agents who
initially favor the unrewarded state toward evidence that would
overturn their initial assessment. Such evidence is harder to obtain,
but it changes the decision whenever it arrives. Agents who favor
the rewarded state continue to search for it, as required by the
incentive constraints. 
The switch sacrifices some decision accuracy while reducing
expected payments. Although the remaining reward rises, the
probability of a paid discovery falls sufficiently to offset
that increase.
Removing one reward prevents a
group of agents from earning it through relatively likely discoveries.
Selective compensation thus emerges from the tradeoff involved in directing
discovery, even for a buyer who values both states equally.

A further increase in acquisition cost produces a discrete reduction
in the optimal reward. With weaker search incentives, failure becomes
less informative, and some agents who strongly favor the rewarded
state generate no improvement in the principal's decision: both a
certificate and silence lead to the same action. Their search still
earns payments, however, because the common reward also induces
valuable discoveries from other agents. Reducing the bonus limits
these payments while preserving useful investigation. This result
highlights a distinction relevant to the procurement of expert-generated data.
In the high-cost regime, a higher prior probability of the rewarded state increases the chance of a verified discovery, but the investigation has no decision value once that probability is sufficiently high.
Evaluating a procurement contract therefore requires accounting for
how the information it generates improves on the knowledge already available through the expert's assessment.

\paragraph{Related literature.}

Our paper connects three strands of literature. The first studies
contracting for information acquisition. \citet{Yoder2022}
characterizes optimal contracts for researchers with privately known
acquisition costs. His analysis allows researchers to choose both
how much and how to learn, and shows how screening distorts
experimentation when realized posterior beliefs are contractible.
\citet{WhitmeyerZhang2023} characterize minimum-cost contracts that
induce the acquisition and truthful reporting of soft information.
The principal and agent initially share the same prior, and payments
can depend on reports and the subsequently observed state.
In our setting, the expert's private initial information affects
both the probability of discovering evidence and its incremental
value for the buyer's decision. Payments depend on certificates
generated through costly search, and the state remains unverifiable
when no certificate is produced.\footnote{Foundational
contributions include \citet{Lambert1986},
\citet{DemskiSappington1987}, and \citet{Osband1989}.
See also \citet{ChadeKovrijnykh2016},
\citet{RappoportSomma2017}, and
\citet{SharmaTsakasVoorneveld2025}.}

The second strand concerns contracting with adverse selection and
moral hazard. \citet{WhitmeyerZhang2026Testing} study whether
outcome-contingent contracts can distinguish experts from
nonexperts when initial beliefs are private and experts can learn.
They identify conditions under which learning makes screening
possible. We instead characterize the nonnegative payments that
optimally procure evidence from a privately informed expert.
The interaction between initial expertise and subsequent search
creates opposing incentives over the direction of investigation
and leads to pooling of discovery bonuses across reports.
\citet{GottliebMoreira2022} provide a complementary connection
through their conditions for single-contract optimality.
Their multiplicative-separability condition does not hold in our
acquisition technology, so our pooling result requires a separate
argument.\footnote{
Our formulation of joint reporting, acquisition, and disclosure
incentives draws on the methodological framework of
\citet{BDL2026}, who provide general conditions for simplifying
evidence-acquisition mechanisms. We establish the reduction
needed for our specific contracting protocol.
Related work studies screening existing knowledge
\citep{Dasgupta2023,DasguptaXia2025}.
\citet{WittbrodtYoder2026} study costly experimentation by a
privately informed agent through menus of observable experiments
without transfers.
}

The third strand concerns the design and sale of information;
see \citet{BergemannBonatti2019} for a survey.
\citet{BergemannBonattiSmolin2018} analyze a seller who designs
and prices experiments for a buyer with private initial information.
Our procurement problem places private information on the supplier's
side: the buyer sets wages to induce costly evidence production,
while the expert privately chooses search direction and effort
under a given acquisition technology.\footnote{
\citet{MakimattilaShangShirakawa2026} study a
certifier selling tests to a privately informed sender.
\citet{BergemannBonattiGan2022} and \citet{QuanSong2026} examine data
markets with informational externalities and endogenous participation,
respectively.
}

\paragraph{Organization.}
Section~\ref{sec:model} presents the model. Section~\ref{sec:bench} develops first-best and
second-best benchmarks with observable expertise. Section~\ref{sec:optimal}
characterizes optimal contracts with private expertise and examines
how acquisition costs shape discovery incentives and decision value.
Section~\ref{sec:con} concludes. All proofs are relegated to the appendices.

\section{Model}\label{sec:model}

There is one principal (he) and one agent (she).
There is an unobserved binary state $\state\in\{0,1\}$.
The agent has private information about the state and can
exert costly effort to acquire further information in the
form of hard evidence.

\paragraph{Expertise.} The agent privately observes an informed prior $\type\in(0,1)$,
where $\Pr(\state=1\mid\type)=\type$.
The principal knows that $\type$ follows a uniform distribution
over $(0,1)$, so his uninformed prior is
$\E[\type]=\tfrac{1}{2}$. 

\paragraph{Discovery.} The agent can acquire perfectly revealing hard evidence about
one of the states. She privately chooses a target state $\hat\state \in \{0,1\}$ and
exerts effort $q\in[0,1]$ at a cost
$c(q)=\tfrac{kq^2}{2}$, where $k>0$.
If she searches state $1$, a certificate of that state arrives
with probability $\type q$; if she searches state $0$, a certificate
of that state arrives with probability $(1-\type)q$.
Otherwise, no evidence is available.
The agent may withhold acquired evidence but cannot fabricate it.

\paragraph{Contracts and payoffs.}
The principal commits to a deterministic, nonnegative wage
contract $w(\hat\type,\cert)$.
The agent first reports her prior $\hat\type\in(0,1)$, then chooses
her search target and effort, and finally discloses evidence
or remains silent.
Here, $\cert\in\{0,1,\emptyset\}$ denotes the principal's
observation: a certificate of state $0$, a certificate of
state $1$, or silence.
Participation is voluntary, and rejecting the contract gives
the agent zero utility.

After observing the report and disclosure, the principal
chooses an action $a\in\{0,1\}$ to match the state.
A mismatch generates a quadratic loss $(a-\state)^2$.
Combined with the wage payment, the principal's ex post
payoff is
\[
u_P(a,\state,\hat\type,\cert)
=-(a-\state)^2-w(\hat\type,\cert).
\]
The agent cares only about her wage and effort cost:
\[
u_A(q,\hat\type,\cert)
=w(\hat\type,\cert)-\frac{kq^2}{2}.
\]

Since the agent does not value the action, choosing an
ex post optimal action does not affect her incentives.
The principal therefore chooses
\[
a(\hat\type,\cert)\in
\arg\min_{a\in\{0,1\}}
\E\bigl[(a-\state)^2\mid\hat\type,\cert\bigr].
\]
The conditional expectation accounts for the acquisition
and disclosure behavior induced by the contract.
A disclosed certificate reveals the state perfectly,
whereas silence may reflect either unsuccessful search
or withheld evidence.

\paragraph{Direct formulation.}
We normalize certificate payments to be no lower than the
silence payment and define the certificate bonuses as
\begin{equation}
\label{FD}
\begin{aligned}
W_0(\hat\type)
&:=w(\hat\type,0)-w(\hat\type,\emptyset)\geq0,\\
W_1(\hat\type)
&:=w(\hat\type,1)-w(\hat\type,\emptyset)\geq0,
\end{aligned}
\qquad
\text{for every }\hat\type\in(0,1).
\tag{FD}
\end{equation}
Because the agent can withhold evidence, this normalization
preserves her reporting and acquisition incentives while
making full disclosure weakly optimal.
Silence then means that no certificate was acquired.

Let $q(\type)$ denote equilibrium effort.
Given \eqref{FD}, incentive compatibility requires the
truthful report and equilibrium effort to maximize the
agent's payoff over all joint deviations in her report,
search target, and effort:
\begin{equation}
\label{IC}
\begin{aligned}
(\type,q(\type))\in
\arg\max_{\substack{\hat\type\in(0,1)\\q\in[0,1]}}
&w(\hat\type,\emptyset)
+q\max\bigl\{
\type W_1(\hat\type),\,
(1-\type)W_0(\hat\type)
\bigr\}-\frac{kq^2}{2},
\end{aligned}
\tag{IC}
\end{equation}
for every $\type\in(0,1)$.
The inner maximum determines the search target by comparing
the expected wage increments from searching the two states.
Thus, after any report, the agent remains free to choose
her own optimal target and effort.

The following lemma establishes that these conditions
provide a lossless formulation of the contracting problem.

\begin{lemma}
\label{lem:directformulation}
Without loss, the principal may restrict attention  to truthful reporting
and full disclosure, for which the feasible set of the contracts is characterized by nonnegativity, 
\eqref{FD}, and  \eqref{IC}.\footnote{
Participation holds automatically because the agent can secure a nonnegative payoff by exerting zero effort.
The argument follows the revelation-and-obedience approach of
\citet{BDL2026}. We provide a direct proof to verify that the
reduction preserves deterministic, nonnegative wage contracts
and is compatible with the principal's ex post optimal action.
}
\end{lemma}

\paragraph{The principal's problem.}
The principal chooses a contract from the feasible set
characterized in Lemma \ref{lem:directformulation} to solve
\begin{equation}
\label{P}
\max_{w\geq0}\;
\E\bigl[
u_P(a(\type,\cert),\state,\type,\cert)
\bigr]
\quad
\text{subject to \eqref{FD} and \eqref{IC}.}
\tag{P}
\end{equation}
The expectation is evaluated at a truthful, full-disclosure
equilibrium and accounts for the induced acquisition behavior.
When the agent is indifferent among acquisition choices,
we select a choice that maximizes the principal's expected
payoff.

\section{Benchmark Analysis: Free Expertise}\label{sec:bench}

This section presents the optimal contracts in two benchmark
environments in which the principal directly observes $\type$.
In the first benchmark (``first best''), the principal also
controls evidence acquisition. In the second benchmark
(``second best''), the agent privately chooses her search
target and effort.

\paragraph{First best.}
The principal observes $\type$, prescribes the search target
and effort, and reimburses the acquisition cost.
Paying $\tfrac{kq^2}{2}$ conditional on compliance leaves
the agent zero rent. This payment does not depend on whether
a certificate arrives, so disclosure is weakly optimal.

For each $\type$, the principal chooses the target state
$\hat\state$ and effort $q$ to maximize
\[
\E\bigl[
-(a(\type,\cert)-\state)^2
\mid\type,\hat\state,q
\bigr]
-\frac{kq^2}{2}.
\]
A certificate perfectly reveals the state and therefore
eliminates decision loss. A mistake can occur only when
no certificate arrives and the action chosen after silence
does not match the state.

For any given effort, searching the less likely state is
optimal.\footnote{
\citet{CheMierendorff2019} likewise show that,
conditional on experimentation, searching for evidence of the
less likely state is optimal in their one-period example.
In their dynamic setting, however, the option to continue
learning can make searching for evidence of the more likely
state optimal. Our one-shot procurement setting provides a
different rationale for such confirmatory search: the interaction
between private expertise (adverse selection) and hidden
evidence acquisition (moral hazard).
}
To see this, suppose $\type\leq\tfrac{1}{2}$.
If the principal searches state $1$, he chooses action $1$
upon receiving its certificate and action $0$ after silence.
The expected loss is $\type(1-q)$: a mistake occurs precisely
when the state is $1$ but no certificate arrives.

Now consider searching state $0$ at the same effort.
The principal chooses action $0$ upon receiving its
certificate. If he chooses action $1$ after silence,
the expected loss is $(1-\type)(1-q)$.
If he instead chooses action $0$ after silence,
he always takes action $0$, giving an expected loss of
$\type$. Thus, the smallest loss from searching state $0$ is $\min\bigl\{\type,(1-\type)(1-q)\bigr\}
\geq \type(1-q)$, 
implied by
$\type\leq\tfrac{1}{2}$.
Searching state $1$ therefore gives a weakly smaller loss.
The argument for $\type>\tfrac{1}{2}$ is symmetric.\footnote{
At $\type=\tfrac{1}{2}$, we select target $1$ and action $0$
after silence in both benchmarks.
}
At $q=1$, either target reveals the state perfectly,
so the target is immaterial.

The minimum expected decision loss at effort $q$ is therefore
$\min\{\type,1-\type\}(1-q)$.
The principal's problem reduces to
\[
\min_{0\leq q\leq1}
\left\{
\min\{\type,1-\type\}(1-q)+\frac{kq^2}{2}
\right\}.
\]
The optimal effort schedule is
\[
q(\type)=
\min\left\{
\frac{\min\{\type,1-\type\}}{k},1
\right\}.
\]
Full effort occurs precisely when $k\leq\tfrac{1}{2}$
and $\type\in[k,1-k]$.
Effort is interior otherwise.

\paragraph{Second best.}
The principal still observes $\type$ and can tailor the
contract to that type, but the agent privately chooses
the target $\hat\state$ and effort $q$.
The principal must now reward evidence to induce acquisition.

Suppose the principal pays only for a state-$1$ certificate,
with zero payments after silence and for the other certificate.
Recall that
$W_1(\type)=w(\type,1)-w(\type,\emptyset)$.
The agent's effort solves
\[
\max_{0\leq q\leq1}
\left\{
\type qW_1(\type)-\frac{kq^2}{2}
\right\},
\]
yielding
\[
q=
\min\left\{
\frac{\type W_1(\type)}{k},1
\right\}.
\]
To induce $0<q<1$, the principal must offer
$W_1(\type)=\tfrac{kq}{\type}$,
giving an expected payment of
$\type qW_1(\type)=kq^2$.
To induce $q=1$, the smallest reward is
$\tfrac{k}{\type}$, again giving an expected payment of
$kq^2=k$.
The calculation for a state-$0$ certificate is symmetric.

A positive silence payment cannot reduce the cost of
inducing effort. Effort depends on the certificate's bonus
above silence, while the silence payment adds to expected
transfers. The same reasoning applies to the chosen target
when both certificates are rewarded.
Thus, the minimum expected payment required to induce
effort $q$ is $kq^2$. Compared with the actual effort cost $\tfrac{1}{2}kq^2$, the additional incentive cost is the agent's rent due to hidden action. 

Because this implementation cost does not depend on the
target state, searching the less likely state remains
optimal. The principal solves
\[
\min_{0\leq q\leq1}
\left\{
\min\{\type,1-\type\}(1-q)+kq^2
\right\}.
\]
The optimal effort and payment for the targeted certificate
are
\[
\begin{aligned}
q(\type)
&=
\min\left\{
\frac{\min\{\type,1-\type\}}{2k},1
\right\},\\
w(\type,\hat\state)
&=
\min\left\{
\frac{1}{2},
\frac{k}{\min\{\type,1-\type\}}
\right\}.
\end{aligned}
\]
The principal pays zero after silence and for the other
certificate. As in the first best, he chooses the more
likely state after silence and matches the state revealed
by a certificate.
When both effort schedules are interior, hidden acquisition
halves effort relative to the first best.
Full effort now occurs precisely when $k\leq\tfrac{1}{4}$
and $\type\in[2k,1-2k]$.

\medskip

By making expertise freely observable, these benchmarks
isolate the cost of hidden acquisition.
The principal must pay an incentive rent in addition to
reimbursing the agent's acquisition cost.
This raises the cost of discovery and reduces effort
whenever second-best effort is interior.

\section{Optimal Contract for Expertise and Discovery}\label{sec:optimal}

We now solve \eqref{P} when expertise is private. The second-best
contract cannot be offered as a menu indexed by the reported prior.
Reports near zero would earn a bonus of $\tfrac{1}{2}$ for state-$1$ evidence,
while reports near one would earn the same bonus for state-$0$
evidence. An agent with access to both options would select the
reward for her more likely state. The difficulty is therefore to
induce discovery while respecting the joint reporting and acquisition
constraint \eqref{IC}.  In what follows, we characterize the exact restrictions and highlight the interplay between hidden expertise and hidden discovery.

\medskip

The first implication of \eqref{IC} concerns the ordering of search
targets and bonuses. A higher prior makes state-$1$ search more
attractive and state-$0$ search less attractive. Within either target,
a larger bonus makes the agent's payoff more responsive to the
probability of that state. 

\begin{lemma}[Monotone sorting of search targets and bonuses]
\label{lem:targetsorting}
Consider any contract satisfying \eqref{FD} and \eqref{IC}. 
As the prior increases from $0$ to $1$, the induced agent's choices are ordered in three groups with any group possibly empty: positive-effort
state-$0$ search, zero effort, and positive-effort state-$1$ search. Within each search group, the selected
certificate bonus is nondecreasing in the probability of the targeted
state.
\end{lemma}The proof exploits a single-crossing property of the agent's
preferences over bonuses and targets. For a fixed report, a higher
prior makes state-$1$ search more attractive and state-$0$ search
less attractive, whereas zero effort yields a payoff
independent of the prior. Within either target, a larger bonus
strengthens the payoff's response to the probability of that state.
These comparisons hold after effort is optimized, including when
effort is capped. Silence payments shift payoff levels without
changing how payoff differences vary with the prior. If two types
selected targets in the reverse order, each could imitate the
other's report and target and then optimize effort. Their two
incentive constraints would contradict single crossing. The same
argument thus orders both search targets and bonuses.

The target ordering rules out the second-best target  assignment, which
searches state $1$ when $\type$ is below $\tfrac{1}{2}$ and state $0$ otherwise. Incentive compatibility therefore limits the
principal's ability to direct discovery toward the less likely
state. It does not, however, rule out such search altogether.
Rewarding only state-$1$ evidence induces search for the less
likely state below $\tfrac{1}{2}$, together with search for the
more likely state above $\tfrac{1}{2}$.

\medskip

We next examine the implications of bonus ordering within a target
group. To do so, we compare the principal's conditional costs under
two bonuses for the same search target. Without loss, consider state-$1$ search. After the agent optimizes effort, the
principal's expected cost, excluding the silence payment, is
\begin{equation}\label{eq:conditional}
 \min\{\type(1-q),1-\type\}+\type W_1q,
 \qquad
 q=\min\left\{\frac{\type W_1}{k},1\right\}.
\end{equation}
The first term is expected decision loss when the principal chooses
optimally after silence; the second is the expected bonus payment.
Because the silence payment is nonnegative, this expression is a
\textit{lower bound} on actual cost. Note that truth-report incentives may require a positive silence payment

\begin{lemma}
\label{lem:bonuscomparison}
For state-$1$ search, fix two bonuses
$0\le W_1<W_1'\le\tfrac{1}{2}$.
If the higher bonus $W_1'$ yields weakly lower cost
in \eqref{eq:conditional} than $W_1$ at some prior, it also does
so at every lower prior. The reflected statement holds for
state-$0$ search, with types ordered by $1-\type$.
\end{lemma}
Lemma~\ref{lem:bonuscomparison} implies that the principal is more
willing to offer a higher bonus when the target state is less
likely. This echoes the benchmarks' preference for directing
discovery toward the less likely state.

The two lemmas reveal opposing orderings within each target group.
Incentive compatibility requires bonuses to be nondecreasing in
the probability of the targeted state, whereas the principal's
relative preference for a higher bonus moves in the opposite
direction. The following proposition establishes that the 
pooling minimizes the lower bound of incentive cost in \eqref{eq:conditional}.

\medskip

\begin{proposition}
\label{prop:bonuspooling}
Fix an interval of priors assigned to search state \(1\). Among all nonnegative bonus schedules that satisfies \eqref{IC} and \eqref{FD}, the incentive cost lower bound in \eqref{eq:conditional} is minimized by a constant bonus in \([0,\tfrac12]\).
The reflected statements hold for state-$0$ search.
\end{proposition}
This reduces the remaining problem to choosing a common bonus for each certificate and a cutoff separating the search targets. 
Theorem \ref{thm:twostructures}
establishes which such choices are jointly incentive compatible and attain the relaxed bound.

\paragraph{Optimal payment structures.}
By pooling the bonuses within each target group, the relaxed problem
reduces to two constant bonuses and a cutoff between search targets.
When a bonus is independent of the report, we omit its argument and
write $W_0$ or $W_1$.

The remaining issue is whether the two pooled bonuses implement
the cutoff used to evaluate the lower bound of incentive cost.
Under report-independent bonuses with $W_0+W_1>0$, the agent searches
state $1$ precisely when $\type W_1\ge(1-\type)W_0$, with either
target optimal at equality. Hence the induced cutoff is
\[
 \bar\type=\frac{W_0}{W_0+W_1}.
\]
Pooling the two groups separately need not satisfy this condition
at a prescribed cutoff. The relaxation allows the cutoff to be
chosen independently; we must then establish that a minimizing
choice can be implemented. In other words, the lower bound of incentive cost in \eqref{eq:conditional} must be attained by our proposed contract.

\begin{theorem}[Optimal contracts for expertise and discovery]
\label{thm:twostructures}
There exist thresholds $0<k_1<k_2<\tfrac{1}{2}$ 
such that for every $k>0$, an optimal contract in \eqref{P} can be
chosen with report-independent certificate payments and zero silence
payment:
\[
 w(\hat\type,0)=W_0,\qquad
 w(\hat\type,1)=W_1,\qquad
 w(\hat\type,\emptyset)=0
 \quad\text{for every report }\hat\type.
\]
The optimal bonuses are characterized as follows.
\begin{enumerate}[label=(\roman*)]
\item {Equal rewards for both certificates.}
For $0<k\le k_1$, set $W_0=W_1$, where $W_1$ is the unique root in
$(k,\min\{2k,\tfrac{1}{2}\})$ of
\begin{equation}\label{eq:thirdsymmetric}
 2W_1^4-(12k+1)W_1^3-4k^3W_1+8k^3=0.
\end{equation}
Each type searches her more likely state.

\item {A reward for one certificate.}
For $k\ge k_1$, set $W_0=0$ and reward only state-$1$ evidence.\footnote{
It is equivalent to use the reflected contract rewarding only state-$0$ evidence.
}
For $k_1\le k\le k_2$, the positive bonus $W_1$ is the unique root
in $(k,\tfrac{1}{2})$ of
\begin{equation}\label{eq:thirdcapped}
 3W_1^3+k^2W_1-2k^2=0.
\end{equation}
All types search state $1$, and types with
$\type\ge\tfrac{k}{W_1}$ exert full effort.

For $k\ge k_2$, the positive bonus $W_1$ can instead be chosen as
the smallest root in $(0,\min\{k,\tfrac{1}{2}\})$ of
\begin{equation}\label{eq:thirdinterior}
 2W_1=(1+\sqrt{1-W_1/k})^{-3}.
\end{equation}
All types then exert interior effort.
\end{enumerate}
At $k=k_1$, both payment structures are optimal. At $k=k_2$, the
two distinct bonuses in part (ii) are both optimal. The thresholds
are the unique cost crossings between these adjacent cases with $k_1\approx0.10, k_2\approx0.28$.
\end{theorem}
The contract shapes discovery along two margins. The extensive
margin concerns which certificates receive rewards and the resulting
search targets; the intensive margin concerns effort conditional
on a target. The choice between payment structures reflects how
these margins jointly determine the ease of inducing discovery
and its value for decisions.
The agent favors the target most likely to yield a reward. The
principal values evidence that changes his action: a certificate of
the less likely state overturns the action favored by the prior,
whereas one of the more likely state confirms it. Searching the
more likely state therefore improves the decision only if effort
makes silence sufficiently informative to reverse that action.

When acquisition is cheap, equal rewards make this possible across
all types. Each type searches her more likely state, whose
probability is at least $\tfrac{1}{2}$, so a small common bonus can
induce intensive search throughout the type distribution. At full
effort, both a certificate and silence reveal the state; near full
effort, the choice of target has little effect on decision accuracy.
Intensive search thus limits the cost of accommodating the agent's
preferred targets under \eqref{IC}.

\begin{figure}[htbp]
\centering
\resizebox{\linewidth}{!}{%
\begin{tikzpicture}[
    x=4cm, y=3.6cm,
    font=\footnotesize,
    line cap=round,
    line join=round,
    optimal/.style={black, line width=1.5pt},
    firstbest/.style={blue!75!black, line width=1pt,
        dash pattern=on 6pt off 4pt},
    secondbest/.style={orange!85!black, line width=1pt,
        dash pattern=on 1pt off 3pt}
]
% Bonuses are the solutions of the equations in Theorem 1.
% (a) k=0.1: W=0.1575390818550989.
\def\RatioA{1.575390818550989}
\def\LeftA{0.365236874415849}
\def\RightA{0.634763125584151}
% (b) k=0.2: W1=0.2838969333644729.
\def\CapB{0.704481015803139}
% (c) k=0.4: W1=0.0723122004405934.
\def\RatioC{0.180780501101484}

% Identical axes and scales. The light grid also marks full effort.
\foreach \PanelShift in {0cm,5.5cm,11cm}{
    \begin{scope}[xshift=\PanelShift]
        \foreach \v/\lab in
            {0.25/{\tfrac14},0.5/{\tfrac12},0.75/{\tfrac34},1/1}{
            \draw[gray!15, line width=0.3pt] (0,\v)--(1,\v);
            \draw[gray!65] (0,\v)--(-0.015,\v)
                node[left,black] {$\lab$};
        }
        \draw[->,gray!70] (0,0)--(1.07,0)
            node[right,black] {$\type$};
        \draw[->,gray!70] (0,0)--(0,1.10)
            node[above,black] {$q$};
        \foreach \u/\lab in {0/0,0.5/{\tfrac12},1/1}{
            \draw[gray!65] (\u,0)--(\u,-0.018)
                node[below,black] {$\lab$};
        }
    \end{scope}
}

% Draw the solid curve first so both dashed benchmarks remain visible
% wherever all three schedules coincide at q=1.
\begin{scope}
    \draw[optimal] (0,1)--(\LeftA,1)
        --(0.5,{\RatioA/2})--(\RightA,1)--(1,1);
    \draw[firstbest] (0,0)--(0.1,1)--(0.9,1)--(1,0);
    \draw[secondbest] (0,0)--(0.2,1)--(0.8,1)--(1,0);
    \node[align=center,anchor=north] at (0.5,-0.23)
        {(a) $k=0.1<k_1$\\[3pt]Equal rewards};
\end{scope}

\begin{scope}[xshift=5.5cm]
    \draw[optimal] (0,0)--(\CapB,1)--(1,1);
    \draw[firstbest] (0,0)--(0.2,1)--(0.8,1)--(1,0);
    \draw[secondbest] (0,0)--(0.4,1)--(0.6,1)--(1,0);
    \node[align=center,anchor=north] at (0.5,-0.23)
        {(b) $k_1<k=0.2<k_2$\\[3pt]State-$1$ reward; capped effort};
\end{scope}

\begin{scope}[xshift=11cm]
    \draw[optimal] (0,0)--(1,\RatioC);
    \draw[firstbest] (0,0)--(0.4,1)--(0.6,1)--(1,0);
    \draw[secondbest] (0,0)--(0.5,0.625)--(1,0);
    \node[align=center,anchor=north] at (0.5,-0.23)
        {(c) $k=0.4>k_2$\\[3pt]State-$1$ reward; interior effort};
\end{scope}

\begin{scope}[x=1cm,y=1cm]
    \draw[optimal] (1.9,-2.2)--(2.6,-2.2)
        node[right] {Optimal contract};
    \draw[firstbest] (6.7,-2.2)--(7.4,-2.2)
        node[right] {First best};
    \draw[secondbest] (10.6,-2.2)--(11.3,-2.2)
        node[right] {Second best};
\end{scope}
\end{tikzpicture}%
}
\caption{Allocation of effort.
\newline {\footnotesize
Panel (a) uses $W_0=W_1\approx0.158$;
panel (b) uses $W_0=0$ and $W_1\approx0.284$;
panel (c) uses $W_0=0$ and $W_1\approx0.072$.
All panels use the same vertical scale.}}
\label{fig:effort}
\end{figure}
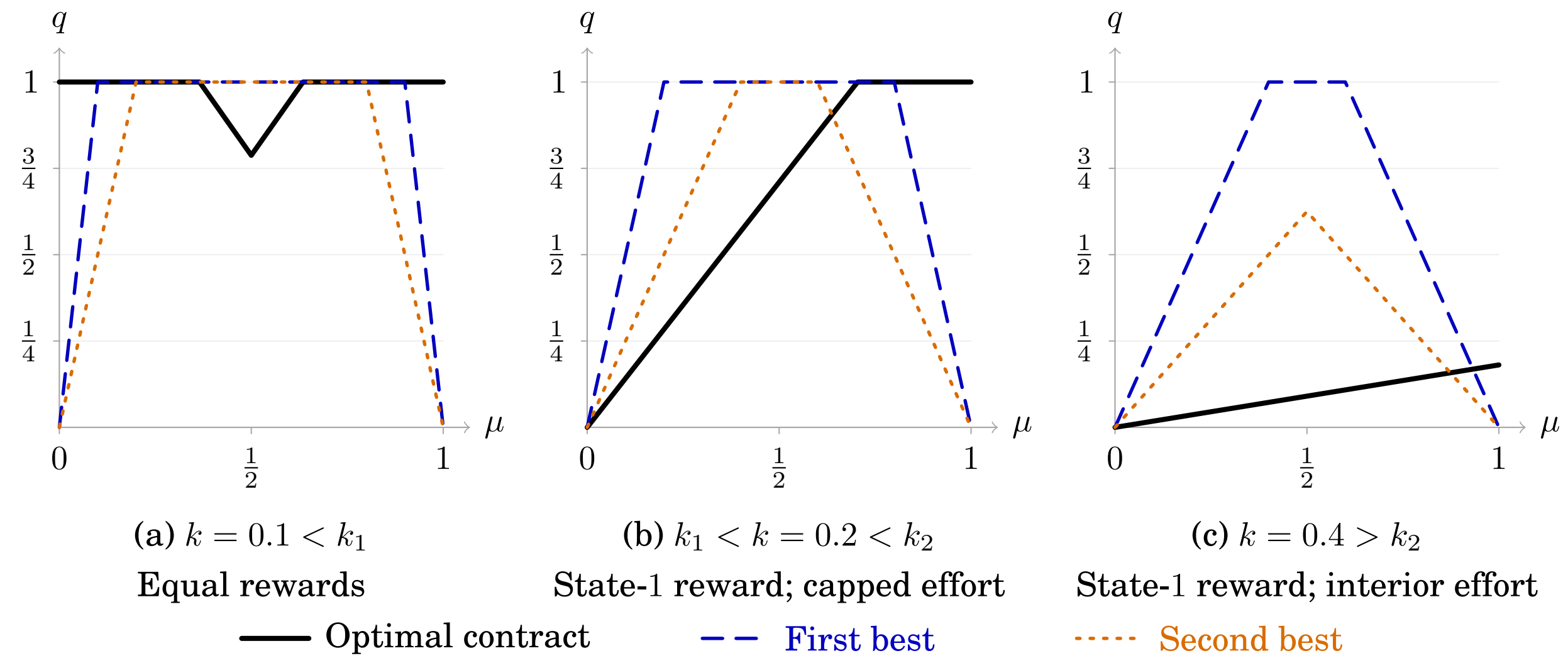

Figure~\ref{fig:effort} compares the allocation of effort under
private expertise with the two observable-expertise benchmarks.
Both benchmarks concentrate effort near $\type=\tfrac{1}{2}$,
where initial uncertainty is greatest; hidden acquisition reduces
its level but preserves this pattern. Under private expertise,
pooled bonuses induce greater effort from types more likely to
discover rewarded evidence. Equal rewards therefore induce the
least effort at the center and full effort at the extremes in
panel (a), whereas rewarding only state-$1$ evidence makes effort
increase with $\type$ in panels (b) and (c). Some extreme types
thus search more intensively than in the first best, although
their search targets also differ.

The transition between the regimes illustrated in panels (b) and
(c) involves a downward jump in effort. As $k$ crosses $k_2$, the
optimal bonus switches from above $k$ to below it, reducing effort
for every type and eliminating the full-effort region at once.
Both benchmark effort schedules vary continuously with $k$.

To illustrate how the optimal contract affects the decision value, we formally define the \textit{value of discovery}.
Given expertise \(\mu\), target \(\hat\theta\), and effort \(q\), the value of discovery is
\[
\mathcal V(\mu,\hat\theta,q)
:=
\min\{\mu,1-\mu\}
-
\mathbb E\!\left[
(a(\mu,e)-\theta)^2
\,\middle|\,\mu,\hat\theta,q
\right],
\]
where acquired evidence is fully disclosed and the principal chooses an optimal action after either a certificate or silence. The first term is the minimum expected loss given expertise alone; the second is the expected loss after discovery. Thus, \(\mathcal V\) measures expected loss avoidance before subtracting payments or acquisition costs.

% Requires \usepackage{tikz,graphicx}

\begin{figure}[htbp]
\centering
\resizebox{\linewidth}{!}{%
\begin{tikzpicture}[
    x=3.8cm, y=6.4cm,
    font=\footnotesize,
    samples=80,
    line cap=round,
    line join=round,
    optimal/.style={black, line width=1.5pt},
    firstbest/.style={
        blue!75!black, line width=1pt,
        dash pattern=on 6pt off 4pt
    },
    secondbest/.style={
        orange!85!black, line width=1pt,
        dash pattern=on 1pt off 3pt
    }
]

% Panel (a): k = 0.1, equal rewards.
\def\RatioA{1.575390818550989}
\def\LeftA{0.365236874415849}
\def\RightA{0.634763125584151}

% Panel (b): k = 0.2, state-1 reward with some full effort.
% W_1 = 0.2838969333644729.
\def\RatioB{1.419484666822364}
\def\CapB{0.704481015803139}

% Panel (c): k = 0.4, state-1 reward with interior effort.
% W_1 = 0.0723122004405934.
\def\RatioC{0.180780501101484}
\def\ZeroC{0.524904775934143}

% Common axes.
\foreach \PanelShift in {0cm,5.2cm,10.4cm}{
    \begin{scope}[xshift=\PanelShift]
        \foreach \v in {0.1,0.2,0.3,0.4,0.5}{
            \draw[gray!65] (0,\v) -- (-0.012,\v);
        }

        \draw[->,gray!70] (0,0) -- (1.06,0)
            node[right,black] {$\type$};
        \draw[->,gray!70] (0,0) -- (0,0.55)
            node[above,black] {$\mathcal V$};

        \foreach \u/\lab in
            {0/0,0.5/{\tfrac{1}{2}},1/1}{
            \draw[gray!65] (\u,0) -- (\u,-0.008)
                node[below,black] {$\lab$};
        }
    \end{scope}
}

% ---------------- Panel (a) ----------------
\begin{scope}
    % Optimal contract.
    \draw[optimal]
        (0,0) -- (\LeftA,\LeftA)
        plot[domain=\LeftA:0.5]
            (\x,{\RatioA*(1-\x)^2-1+2*\x})
        plot[domain=0.5:\RightA]
            (\x,{\RatioA*\x^2-2*\x+1})
        -- (1,0);

    % First best.
    \draw[firstbest]
        plot[domain=0:0.1]
            (\x,{10*\x^2})
        -- (0.5,0.5) -- (0.9,0.1)
        plot[domain=0.9:1]
            (\x,{10*(1-\x)^2});

    % Second best.
    \draw[secondbest]
        plot[domain=0:0.2]
            (\x,{5*\x^2})
        -- (0.5,0.5) -- (0.8,0.2)
        plot[domain=0.8:1]
            (\x,{5*(1-\x)^2});

    \node[align=center] at (0.5,-0.18)
        {(a) $k=0.1<k_1$\\[2pt]
         Equal rewards};
\end{scope}

% ---------------- Panel (b) ----------------
\begin{scope}[xshift=5.2cm]
    % Optimal contract: all types search state 1.
    % Full effort begins at CapB.
    \draw[optimal]
        plot[domain=0:0.5]
            (\x,{\RatioB*\x^2})
        plot[domain=0.5:\CapB]
            (\x,{\RatioB*\x^2-2*\x+1})
        -- (1,0);

    % First best: full effort on [0.2,0.8].
    \draw[firstbest]
        plot[domain=0:0.2]
            (\x,{5*\x^2})
        -- (0.5,0.5) -- (0.8,0.2)
        plot[domain=0.8:1]
            (\x,{5*(1-\x)^2});

    % Second best: full effort on [0.4,0.6].
    \draw[secondbest]
        plot[domain=0:0.4]
            (\x,{2.5*\x^2})
        -- (0.5,0.5) -- (0.6,0.4)
        plot[domain=0.6:1]
            (\x,{2.5*(1-\x)^2});

    \node[align=center] at (0.5,-0.18)
        {(b) $k_1<k=0.2<k_2$\\[2pt]
         State-$1$ reward};
\end{scope}

% ---------------- Panel (c) ----------------
\begin{scope}[xshift=10.4cm]
    % Optimal contract: all effort is interior.
    % Discovery value vanishes above ZeroC.
    \draw[optimal]
        plot[domain=0:0.5]
            (\x,{\RatioC*\x^2})
        plot[domain=0.5:\ZeroC]
            (\x,{\RatioC*\x^2-2*\x+1})
        -- (1,0);

    % First best.
    \draw[firstbest]
        plot[domain=0:0.4]
            (\x,{2.5*\x^2})
        -- (0.5,0.5) -- (0.6,0.4)
        plot[domain=0.6:1]
            (\x,{2.5*(1-\x)^2});

    % Second best.
    \draw[secondbest]
        plot[domain=0:0.5]
            (\x,{1.25*\x^2})
        plot[domain=0.5:1]
            (\x,{1.25*(1-\x)^2});

    \node[align=center] at (0.5,-0.18)
        {(c) $k=0.4>k_2$\\[2pt]
         State-$1$ reward};
\end{scope}

% Shared legend; coordinates measured in centimeters.
\begin{scope}[x=1cm,y=1cm]
    \draw[optimal] (1.3,-2) -- (2.0,-2)
        node[right] {Optimal contract};
    \draw[firstbest] (6.0,-2) -- (6.7,-2)
        node[right] {First best};
    \draw[secondbest] (10.4,-2) -- (11.1,-2)
        node[right] {Second best};
\end{scope}

\end{tikzpicture}%
}
\caption{Value of discovery.
\newline {\footnotesize
Panel (a) uses $W_0=W_1\approx0.16$;
panel (b) uses $W_0=0$ and $W_1\approx0.284$;
panel (c) uses $W_0=0$ and $W_1\approx0.072$.
All panels use the same vertical scale.
These gross informational benefits exclude payments and acquisition
costs and therefore need not follow the ranking of net payoffs.}
}
\label{fig:discovery-value}
\end{figure}
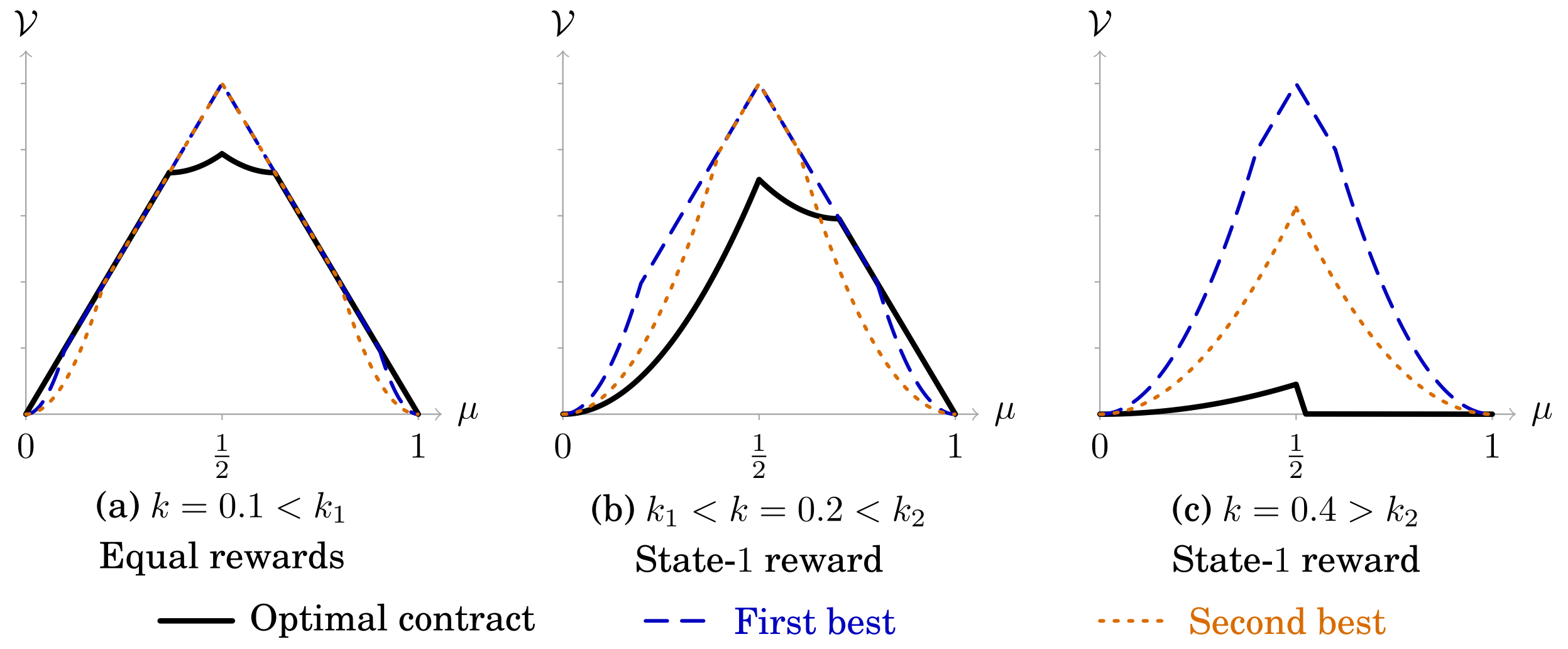

Figure~\ref{fig:discovery-value} shows how the gains from discovery
are distributed across types. In panel (a), equal rewards induce
full revelation near the endpoints but only partial revelation
near $\tfrac{1}{2}$. Nevertheless, discovery value peaks at
$\tfrac{1}{2}$, where expertise alone leaves the most uncertainty.
Near the endpoints, the optimal contract generates greater loss
avoidance than either benchmark, although the absolute gains are
small. This reflects full acquisition where the benchmarks
economize on its cost; the higher curve therefore does not imply
a higher net payoff.

Panel (b) reveals an asymmetry in the gains from discovery.
Types with priors equally far from $\tfrac{1}{2}$ face the same
initial decision loss, but discovery creates more value on the
side favoring the rewarded state. These types exert greater effort
under the common reward, which here more than offsets their search
for the more likely state. Below $\tfrac{1}{2}$, discovery value
rises with $\type$ because both effort and the probability of the
targeted state increase. Above $\tfrac{1}{2}$, greater effort
partly offsets the declining uncertainty left by expertise,
producing the flatter portion of the curve. At
$\type=\tfrac{k}{W_1}\approx0.704$, effort reaches one and
discovery eliminates all remaining decision loss. Its value
thereafter equals $1-\type$: further increases in the prior reduce
the loss that can be avoided without improving the already fully
revealing search. This accounts for the kink in the curve.

Panel (c) shows a sharper divergence between acquisition and
decision value. Although effort increases throughout the type
space, discovery value vanishes above $\type\approx0.525$.
For these types, either a certificate or silence leads to action
$1$. A greater probability of earning the reward therefore
generates additional payments without improving the decision.
In contrast to panel (b), the upper part of the type distribution
contributes no discovery value despite continued acquisition.

\medskip

The first threshold, $k_1$, marks an adjustment along the extensive
margin: the principal switches from rewarding both certificates
to rewarding only one. This change redirects search targets and
also alters effort incentives.
Rewarding state $1$ redirects types below $\tfrac{1}{2}$ toward their less likely state. These
discoveries are harder to induce, but change the decision whenever
they occur, so even modest search is valuable. The same reward also
induces higher-prior types to search state $1$, consistent with the
target ordering required by \eqref{IC}. At $k_1$, with the bonus
optimized under each payment structure, the lower expected payments
under the one-certificate contract exactly offset its greater
decision loss. As $k$ passes this threshold, the payment savings
begin to dominate. The transition changes search targets while
retaining full effort for some types.

The second threshold, $k_2$, changes search intensity within the
one-certificate contract. With the larger bonus, $W_1>k$, silence
induces action $0$ for every report, overturning even strong priors
in favor of state $1$. With the smaller bonus, $W_1<k$, all effort
is interior, and sufficiently high priors continue to induce action
$1$ after silence. For these types, acquisition no longer changes
the decision but still generates payments. The smaller bonus limits
these payments while preserving useful discoveries for lower priors.

The switch is discontinuous because, for $W_1<k$, increasing the
bonus both raises effort and expands the set of priors for which
silence changes the action. These effects make the principal's cost
nonconvex, with two distinct local minima near $k_2$. The intermediate
bonus $W_1=k$, at which full effort would disappear continuously,
is never optimal near $k_2$: a small increase in the bonus still
lowers total cost. At $k_2$, the minima on either side of this
boundary have equal costs. The cost of the larger-bonus contract
then rises faster with $k$, so the optimal bonus jumps below $k$.

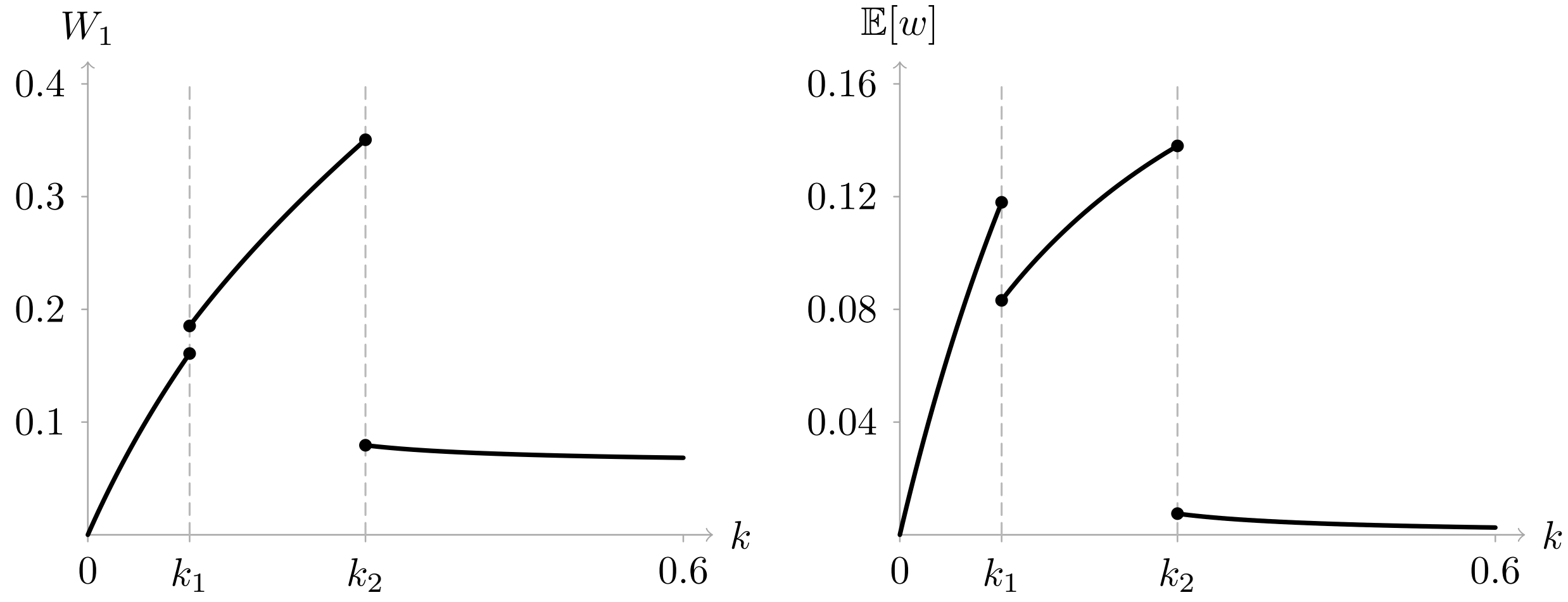
\begin{figure}[htbp]
\centering
\resizebox{\linewidth}{!}{%
\begin{tikzpicture}[
    x=8.8cm, y=10cm,
    font=\footnotesize,
    samples=120,
    line cap=round,
    line join=round,
    optimal/.style={black, line width=1.15pt},
    threshold/.style={gray!55, densely dashed, line width=0.5pt},
    declare function={
        wEqual(\t)=(8*\t^3-1)/(2*(2*\t^3+6*\t-1));
        kEqual(\t)=\t*wEqual(\t);
        pEqual(\t)=wEqual(\t)*(1-\t^2/3-1/(12*\t));
        wCapped(\t)=2*\t^2/(3+\t^2);
        kCapped(\t)=\t*wCapped(\t);
        pCapped(\t)=\t^2*(3-\t^2)/(3*(3+\t^2));
        wInterior(\t)=\t^3/2;
        kInterior(\t)=\t^5/(2*(2*\t-1));
        pInterior(\t)=\t*(2*\t-1)/6;
    }
]

\def\FirstThreshold{0.102571318520453}
\def\SecondThreshold{0.279756601639733}

% Parameters at the branch endpoints.
\def\EqualEnd{0.637858939176147}
\def\CappedStart{0.553497057167160}
\def\CappedEnd{0.798320146798805}
\def\InteriorStart{0.541671571812698}
\def\InteriorEnd{0.515110986690127}

% ---------------- Panel (a): certificate bonus ----------------
\begin{scope}
    \foreach \v in {0.1,0.2,0.3,0.4}{
        \draw[gray!65] (0,\v) -- (-0.006,\v)
            node[left,black] {$\v$};
    }
    \draw[->,gray!70] (0,0) -- (0.63,0)
        node[right,black] {$k$};
    \draw[->,gray!70] (0,0) -- (0,0.42)
        node[above,black] {$W_1$};

    \foreach \u/\lab in
        {0/0,\FirstThreshold/{k_1},\SecondThreshold/{k_2},0.6/{0.6}}{
        \draw[gray!65] (\u,0) -- (\u,-0.006)
            node[below,black] {$\lab$};
    }
    \foreach \u in {\FirstThreshold,\SecondThreshold}{
        \draw[threshold] (\u,0) -- (\u,0.4);
    }

    \draw[optimal] plot[variable=\t,domain=0.5:\EqualEnd]
        ({kEqual(\t)},{wEqual(\t)});
    \draw[optimal] plot[variable=\t,domain=\CappedStart:\CappedEnd]
        ({kCapped(\t)},{wCapped(\t)});
    \draw[optimal] plot[variable=\t,domain=\InteriorEnd:\InteriorStart]
        ({kInterior(\t)},{wInterior(\t)});

    % Both marked bonuses are optimal at each threshold.
    \foreach \u/\v in {
        \FirstThreshold/0.160805645607058,
        \FirstThreshold/0.185315020544862,
        \SecondThreshold/0.350431594093589,
        \SecondThreshold/0.079465411109570}{
        \fill[black] (\u,\v) circle[radius=1.6pt];
    }
    \node[align=center] at ([yshift=-1cm]0.3,0)
        {(a) Optimal certificate bonus};
\end{scope}

% ---------------- Panel (b): expected payment ----------------
% A separate labeled vertical scale keeps this curve readable.
\begin{scope}[xshift=7.2cm,y=25cm]
    \foreach \v/\lab in {0.04/{0.04},0.08/{0.08},0.12/{0.12},0.16/{0.16}}{
        \draw[gray!65] (0,\v) -- (-0.006,\v)
            node[left,black] {$\lab$};
    }
    \draw[->,gray!70] (0,0) -- (0.63,0)
        node[right,black] {$k$};
    \draw[->,gray!70] (0,0) -- (0,0.168)
        node[above,black] {$\mathbb{E}[w]$};

    \foreach \u/\lab in
        {0/0,\FirstThreshold/{k_1},\SecondThreshold/{k_2},0.6/{0.6}}{
        \draw[gray!65] (\u,0) -- (\u,-0.0024)
            node[below,black] {$\lab$};
    }
    \foreach \u in {\FirstThreshold,\SecondThreshold}{
        \draw[threshold] (\u,0) -- (\u,0.16);
    }

    \draw[optimal] plot[variable=\t,domain=0.5:\EqualEnd]
        ({kEqual(\t)},{pEqual(\t)});
    \draw[optimal] plot[variable=\t,domain=\CappedStart:\CappedEnd]
        ({kCapped(\t)},{pCapped(\t)});
    \draw[optimal] plot[variable=\t,domain=\InteriorEnd:\InteriorStart]
        ({kInterior(\t)},{pInterior(\t)});

    \foreach \u/\v in {
        \FirstThreshold/0.117988450976544,
        \FirstThreshold/0.083195356447293,
        \SecondThreshold/0.137993241831967,
        \SecondThreshold/0.007524101934563}{
        \fill[black] (\u,\v) circle[radius=1.6pt];
    }
    \node[align=center] at ([yshift=-1cm]0.3,0)
        {(b) Incentive cost (expected payment)};
\end{scope}

\end{tikzpicture}%
}
\caption{Optimal certificate bonus and incentive cost.
\newline {\footnotesize
Expected payment is averaged over the agent's type and the
discovery outcome. For $k>k_1$, the figure selects the optimal
contract rewarding state-$1$ evidence.
At each threshold, both marked contracts are optimal.}}
\label{fig:bonus-payment}
\end{figure}

Figure~\ref{fig:bonus-payment} distinguishes the reward per
certificate from the expected cost of providing incentives.
At $k_1$, the switch to one rewarded certificate changes the
composition of search. Under equal rewards, every type searches
her more likely state, and many already exert full effort; a
higher bonus increases payments to these types without improving
their search. Removing the state-$0$ reward redirects lower-prior
types toward harder-to-obtain state-$1$ evidence. Their effort
remains responsive to the bonus, and their certificates overturn
the action favored by the prior. This reallocation makes a larger
state-$1$ reward worthwhile. Nevertheless, the aggregate probability
of a paid discovery falls as low-prior types lose access to easily
obtained state-$0$ certificates. At the threshold, the bonus rises
from approximately $0.161$ to $0.185$, while the probability of
payment falls from approximately $0.734$ to $0.449$. The decline
in payment frequency more than offsets the larger reward.

Above $k_2$, the optimal bonus decreases because additional search
improves decisions for a shrinking range of priors. At a given
bonus, higher acquisition costs reduce effort and make silence
less informative. For some types above $\tfrac{1}{2}$, silence
therefore ceases to overturn the action favored by the prior.
Raising the common bonus still increases payments across all
types, including those whose search does not change the decision,
but its informational benefit is concentrated among fewer types.
The marginal benefit of stronger incentives consequently falls
relative to their marginal payment cost, making a smaller bonus
optimal.

Expected payments then fall for two reinforcing reasons: the
principal pays less per certificate, and certificates arrive less
often because both the smaller bonus and the higher acquisition
cost reduce effort. Since all effort is interior in this regime,
\[
 \mathbb{E}[w]
 =\int_0^1 \type W_1\frac{\type W_1}{k}\,d\type
 =\frac{W_1^2}{3k}.
\]
Thus the decline in the bonus compounds the direct effect of
higher costs on payment frequency. As $k$ becomes arbitrarily
large, the bonus approaches $\tfrac{1}{16}$, while expected
payments approach zero: a positive reward supports increasingly
rare discoveries.

\medskip

Theorem~\ref{thm:twostructures} establishes that these contracts
attain the minimum lower bound of incentive cost.
It proves the
existence of an optimal contract of the stated form without requiring
every optimal menu to have that form. Truthful reporting is weakly
optimal under report-independent payments, as allowed by the model's
equilibrium selection. Appendix~\ref{app:theorem} derives the bonuses,
establishes the thresholds, and verifies attainment.

\section{Conclusion}\label{sec:con}

This paper studies how a data buyer procures evidence from an expert
whose initial knowledge and subsequent investigation are private.
Expertise affects both the likelihood of discovery and its value
for the buyer's decision. An optimal contract balances these
considerations through report-independent rewards: equal rewards
for both certificates when acquisition is cheap, and a reward for
only one certificate when it is more costly. At sufficiently high
costs, some rewarded investigation never changes the buyer's action.
These findings highlight the importance of evaluating expert-generated
data by its contribution beyond the expert's initial assessment.

We establish the characterization in a binary-state, binary-action
model with perfectly revealing evidence, uniformly distributed
informed priors, and quadratic effort costs. 
While uniform prior distribution and quadratic effort costs are not essential for our qualitative results, the binary-state, binary-action structure is crucial for the clear contrast between the principal's and the agent's preferred sorting of the search target. Relaxing this assumption is an interesting future research direction. 
  Another important
direction is to allow experts to care about the buyer's decision.
Such preferences would create additional incentives to pursue or
conceal particular findings, linking the design of discovery rewards
to selective disclosure and the buyer's ability to commit to a
decision rule.

\bibliographystyle{ecta}
\bibliography{references}

@article{BDL2026,
  author  = {Ben-Porath, Elchanan and Dekel, Eddie and Lipman, Barton L.},
  title   = {Mechanism Design for Acquisition of/Stochastic Evidence},
  journal = {Theoretical Economics},
  year    = {2026},
  volume  = {21},
  number  = {1},
  pages   = {99--131},
  doi     = {10.3982/TE5974}
}

@misc{WhitmeyerZhang2023,
  author = {Whitmeyer, Mark and Zhang, Kun},
  title  = {Buying Opinions},
  year   = {2023},
  howpublished = {Working paper, \url{https://arxiv.org/abs/2202.05249}},
  note   = {July 31, 2023. arXiv:2202.05249v5}
}

@article{Lambert1986,
  author  = {Lambert, Richard A.},
  title   = {Executive Effort and Selection of Risky Projects},
  journal = {The RAND Journal of Economics},
  year    = {1986},
  volume  = {17},
  number  = {1},
  pages   = {77--88},
  doi     = {10.2307/2555629}
}

@article{DemskiSappington1987,
  author  = {Demski, Joel S. and Sappington, David E. M.},
  title   = {Delegated Expertise},
  journal = {Journal of Accounting Research},
  year    = {1987},
  volume  = {25},
  number  = {1},
  pages   = {68--89},
  doi     = {10.2307/2491259}
}

@article{Osband1989,
  author  = {Osband, Kent},
  title   = {Optimal Forecasting Incentives},
  journal = {Journal of Political Economy},
  year    = {1989},
  volume  = {97},
  number  = {5},
  pages   = {1091--1112},
  doi     = {10.1086/261644}
}

@misc{RappoportSomma2017,
  author = {Rappoport, Daniel and Somma, Valentin},
  title  = {Incentivizing Information Design},
  year   = {2017},
  howpublished = {Working paper, \url{https://ssrn.com/abstract=3001416}},
  doi    = {10.2139/ssrn.3001416}
}

@article{SharmaTsakasVoorneveld2025,
  author  = {Sharma, Salil and Tsakas, Elias and Voorneveld, Mark},
  title   = {Procuring Unverifiable Information},
  journal = {Mathematics of Operations Research},
  year    = {2025},
  volume  = {50},
  number  = {2},
  pages   = {1433--1453},
  doi     = {10.1287/moor.2022.0085}
}

@article{ChadeKovrijnykh2016,
  author  = {Chade, Hector and Kovrijnykh, Natalia},
  title   = {Delegated Information Acquisition with Moral Hazard},
  journal = {Journal of Economic Theory},
  year    = {2016},
  volume  = {162},
  pages   = {55--92},
  doi     = {10.1016/j.jet.2015.12.003}
}

@article{Yoder2022,
  author  = {Yoder, Nathan},
  title   = {Designing Incentives for Heterogeneous Researchers},
  journal = {Journal of Political Economy},
  year    = {2022},
  volume  = {130},
  number  = {8},
  pages   = {2018--2054},
  doi     = {10.1086/720072}
}

@article{GottliebMoreira2022,
  author  = {Gottlieb, Daniel and Moreira, Humberto},
  title   = {Simple Contracts with Adverse Selection and Moral Hazard},
  journal = {Theoretical Economics},
  year    = {2022},
  volume  = {17},
  number  = {3},
  pages   = {1357--1401},
  doi     = {10.3982/TE2992},
  url     = {https://doi.org/10.3982/TE2992}
}

@misc{EpochAI2026FrontierMath,
  author = {{Epoch AI}},
  title  = {About {FrontierMath}},
  year   = {2026},
  url    = {https://epoch.ai/frontiermath/tiers-1-4/about},
  note   = {Accessed October 19, 2026}
}

@article{BergemannBonatti2019,
  author  = {Bergemann, Dirk and Bonatti, Alessandro},
  title   = {Markets for Information: An Introduction},
  journal = {Annual Review of Economics},
  year    = {2019},
  volume  = {11},
  pages   = {85--107},
  doi     = {10.1146/annurev-economics-080315-015439}
}

@article{CheMierendorff2019,
  author  = {Che, Yeon-Koo and Mierendorff, Konrad},
  title   = {Optimal Dynamic Allocation of Attention},
  journal = {American Economic Review},
  year    = {2019},
  volume  = {109},
  number  = {8},
  pages   = {2993--3029},
  doi     = {10.1257/aer.20171000}
}

@misc{WittbrodtYoder2026,
  author = {Wittbrodt, Cole and Yoder, Nathan},
  title  = {Delegating Experiments},
  year   = {2026},
  howpublished = {Working paper, \url{https://ndyoder.github.io/delegating_experiments_063026.pdf}}
}

@misc{PremiseGeospatialCaseStudy,
  author = {{Premise}},
  title  = {Geospatial Solutions {NGO} Uses {Premise}
            to Fill in Location Data Gaps in Ministry Records},
  year   = {n.d.},
  howpublished = {Case study},
  url = {https://premise.com/case-studies/geospatial-solutions-ngo-uses-premise-to-fill-in-location-data-gaps-in-ministry-records/},
  note   = {Accessed October 19, 2026}
}

@misc{Dasgupta2023,
  author = {Dasgupta, Sulagna},
  title  = {Screening Knowledge},
  year   = {2023},
  howpublished = {Working paper, \url{https://ssrn.com/abstract=4403119}},
  note   = {November 8, 2023},
  doi    = {10.2139/ssrn.4403119}
}

@misc{DasguptaXia2025,
  author = {Dasgupta, Sulagna and Xia, Zizhe},
  title  = {Screening Knowledge with Verifiable Evidence},
  year   = {2025},
  howpublished = {Working paper, \url{https://www.economics.utoronto.ca/conferences/index.php/cetc/2025/paper/viewFile/1926/508}}
}

@article{MakimattilaShangShirakawa2026,
  author  = {M{\"a}kimattila, Mikael and Shang, Yucheng and Shirakawa, Ryo},
  title   = {The Design and Price of Certification},
  journal = {Theoretical Economics},
  year    = {2026},
  note    = {Forthcoming},
  url     = {https://econtheory.org/ojs/index.php/te/article/viewForthcomingFile/6640/46344/1}
}

@article{BergemannBonattiSmolin2018,
  author  = {Bergemann, Dirk and Bonatti, Alessandro and Smolin, Alex},
  title   = {The Design and Price of Information},
  journal = {American Economic Review},
  year    = {2018},
  volume  = {108},
  number  = {1},
  pages   = {1--48},
  doi     = {10.1257/aer.20161079}
}

@article{BergemannBonattiGan2022,
  author  = {Bergemann, Dirk and Bonatti, Alessandro and Gan, Tan},
  title   = {The Economics of Social Data},
  journal = {The RAND Journal of Economics},
  year    = {2022},
  volume  = {53},
  number  = {2},
  pages   = {263--296},
  doi     = {10.1111/1756-2171.12407}
}

@misc{QuanSong2026,
  author = {Quan, Li and Song, Yangbo},
  title  = {Learning-Optimal Design of Data Markets},
  year   = {2026},
  howpublished = {Working paper, \url{https://archiquan.github.io/papers/Learning-Optimal-Data-Markets.pdf}},
  note   = {June 25, 2026}
}

@misc{WhitmeyerZhang2026Testing,
  author = {Whitmeyer, Mark and Zhang, Kun},
  title  = {Testing Experts Who Can Learn},
  year   = {2026},
  howpublished = {Working paper, \url{https://arxiv.org/abs/2303.15723v4}},
  note   = {September 3, 2026. arXiv:2303.15723v4},
  doi    = {10.48550/arXiv.2303.15723}
}

@misc{OpenAI2024SWEbenchVerified,
  author = {{OpenAI}},
  title  = {Introducing {SWE-bench Verified}},
  year   = {2024},
  month  = aug,
  howpublished = {\url{https://openai.com/index/introducing-swe-bench-verified/}},
  note   = {Accessed October 18, 2026}
}

\appendix

\section{Proof of Lemma \ref{lem:directformulation}}
Fix any contract $\bar w$ and its induced agent's best response.
Let $r(\type)$ denote type $\type$'s selected equilibrium
report. Define
\[
\begin{aligned}
\widetilde w(\hat\type,\emptyset)
&=\bar w(r(\hat\type),\emptyset),\\
\widetilde w(\hat\type,j)
&=\max\{\bar w(r(\hat\type),j),
         \bar w(r(\hat\type),\emptyset)\},
\qquad j\in\{0,1\}.
\end{aligned}
\]
Since the agent could already withhold a certificate, this
construction preserves her optimized payoff for every
report, target, and effort after relabeling reports.
Any joint deviation under $\widetilde w$ can therefore be
replicated under $\bar w$ by reporting $r(\hat\type)$ and
making the same acquisition choices.
Truthful reporting and the selected equilibrium acquisition
choices consequently remain optimal, while \eqref{FD}
makes full disclosure weakly optimal.

Payments are unchanged at these choices, and observing the
true type and acquired certificate weakly improves the
principal's decision. If the original best response mixes,
select, for each type, among her optimal reports and
acquisition choices one whose conditional cost under this
additional information is no greater than their average.
The restriction therefore entails no loss.

Finally, under \eqref{FD}, optimizing over the target and
disclosure yields exactly the objective in \eqref{IC}.
Thus \eqref{IC} is necessary and sufficient for truthful
reporting and the specified effort to be optimal against
all joint deviations. Zero effort guarantees a nonnegative
silence payment, so participation holds automatically.
\hfill \qedsymbol
\section{Proof of the payment-structure reduction}
\label{app:reduction}

This appendix proves Lemmas~\ref{lem:targetsorting} and
\ref{lem:bonuscomparison}, Proposition~\ref{prop:bonuspooling}, and
the remaining cutoff result used in Theorem~\ref{thm:twostructures}.
Throughout, bonuses are those
in \eqref{FD}, and reporting and acquisition incentives are governed
by \eqref{IC}. For a constant state-$1$ bonus, we write $W_1$;
state-$0$ calculations follow by replacing $\type$ with $1-\type$.

\subsection{Proof of Lemma \ref{lem:targetsorting}}
For this proof, associate each type's selected option with the signed
bonus
\[
 x(\type)=
 \begin{cases}
 -W_0(\type),&\text{positive effort targeting state }0,\\
 0,&q(\type)=0,\\
 W_1(\type),&\text{positive effort targeting state }1.
 \end{cases}
\]
Positive effort requires a positive bonus for the selected target.
Zero effort requires both bonuses at the selected report to be zero,
since otherwise sufficiently small positive effort would be profitable.

For a fixed report and its selected target, let $V(\type,x)$ denote
the acquisition payoff after optimizing effort, excluding the silence
payment, where $x$ is the signed bonus of that option. For $x=0$,
this payoff is zero. Optimizing effort and differentiating gives
\[
 \frac{\partial V(\type,x)}{\partial\type}=
 \begin{cases}
 x\min\left\{\dfrac{(1-\type)(-x)}{k},1\right\},&x<0,\\[4pt]
 x\min\left\{\dfrac{\type x}{k},1\right\},&x\ge0.
 \end{cases}
\]
For every $\type\in(0,1)$, this derivative is strictly increasing in
$x$, including when effort is capped. Hence, for any fixed $x'>x$,
the utility difference $V(\type,x')-V(\type,x)$ strictly increases
with $\type$.

Suppose $\type<\type'$ but $x(\type)>x(\type')$. Each type can copy
the other's report and selected target and then optimize her own
effort. Adding the two incentive constraints implied by \eqref{IC}
cancels silence payments and gives
\[
 V(\type,x(\type))-V(\type,x(\type'))
 \ge
 V(\type',x(\type))-V(\type',x(\type')),
\]
contradicting the strict increase just established. Thus $x(\type)$
is nondecreasing. Its sign orders state-$0$ search, zero effort, and
state-$1$ search. Within the two search groups, respectively,
$W_0(\type)$ is nonincreasing in $\type$ and $W_1(\type)$ is
nondecreasing in $\type$, proving the bonus ordering.

\hfill\qedsymbol

\subsection{Proofs of Lemma \ref{lem:bonuscomparison} and Proposition \ref{prop:bonuspooling}}

Throughout this subsection, let $C(\type,W_1)$ denote the typewise
lower bound of incentive cost in \eqref{eq:conditional}. This
criterion includes both expected decision loss and expected bonus
payments. For an interval of priors, we evaluate the lower bound of
incentive cost by integrating this expression over that interval.

\begin{proof}[Proof of Lemma~\ref{lem:bonuscomparison}]
Fix $0\le W_1<W_1'\le\tfrac{1}{2}$. Define, within this proof,
$h(\type)=\tfrac{2\type-1}{\type^2}$, which increases on $(0,1)$, and
write $[x]_+=\max\{x,0\}$. If neither effort is capped,
\begin{align*}
 \frac{C(\type,W_1')-C(\type,W_1)}{\type^2}
 &=\frac{W_1'^2-W_1^2}{k}\\
 &\quad-\left\{
 \left[\frac{W_1'}{k}-h(\type)\right]_+
 -\left[\frac{W_1}{k}-h(\type)\right]_+
 \right\}.
\end{align*}
The term in braces weakly decreases with $h(\type)$, so the
normalized cost difference is nondecreasing in $\type$.

If only the higher bonus induces capped effort, the same difference
is
\[
 \max\left\{
 \frac{W_1'-1}{\type}+\frac{W_1(1-W_1)}{k},\;
 \frac{1+W_1'}{\type}-\frac{1}{\type^2}-\frac{W_1^2}{k}
 \right\}.
\]
Both entries increase with $\type$, since their derivatives are
$\tfrac{1-W_1'}{\type^2}>0$ and
$\tfrac{2-(1+W_1')\type}{\type^3}>0$.
Once both efforts are capped, the unnormalized cost difference is
$\type(W_1'-W_1)>0$. The regimes occur in this order as $\type$
increases, with any absent regime simply omitted. Continuity at the
boundaries therefore implies that if the higher-minus-lower
difference is nonpositive at some prior, it is nonpositive at every
lower prior. This proves the claim, including weak preferences at
indifference. Replacing $\type$ with $1-\type$ gives the reflected
statement for state-$0$ search.
\end{proof}

\begin{proof}[Proof of Proposition~\ref{prop:bonuspooling}]
Fix an interval $[a,b]$ of types assigned to state-$1$ search.
Lemma~\ref{lem:targetsorting} requires the selected bonus schedule
to be nondecreasing in $\type$. We prove the stronger comparison
that every measurable, nonnegative, nondecreasing schedule has a
lower bound of incentive cost at least as large as that of some
constant bonus in $[0,\tfrac{1}{2}]$. The comparison allows the bonus
schedule to vary without imposing incentive constraints involving
the other target group.

\smallskip
\noindent{\it Step 1: truncate bonuses above $\tfrac{1}{2}$ in the relaxation.}
While effort is uncapped,
\[
 C(\type,W_1)=\min\left\{
 \type+\frac{\type^2}{k}(W_1^2-W_1),\;
 1-\type+\frac{\type^2W_1^2}{k}\right\}.
\]
Both entries are nondecreasing in $W_1\ge\tfrac{1}{2}$. Once effort is capped,
$C(\type,W_1)=\type W_1$, which is increasing. Continuity at the cap
therefore gives
\[
 C\left(\type,\min\left\{W_1,\frac{1}{2}\right\}\right)
 \le C(\type,W_1).
\]
Truncation preserves monotonicity and weakly reduces the lower
bound of incentive cost.

\smallskip
\noindent{\it Step 2: move bonus cutoffs to extreme points.}
Consider a finite nondecreasing schedule on $[a,b]$,
with distinct bonus levels
$0\le W_1^{(1)}<\cdots<W_1^{(m)}\le\tfrac{1}{2}$ and ordered cutoffs
$a=t_0\le t_1\le\cdots\le t_m=b$.
Holding the bonus levels fixed, its lower bound of incentive cost is
\[
 \sum_{i=1}^m\int_{t_{i-1}}^{t_i}C(\type,W_1^{(i)})\,d\type.
\]
For two adjacent nonempty blocks, hold their union fixed and move
their common cutoff $t$. The derivative of this bound is
\[
 C(t,W_1^{(i)})-C(t,W_1^{(i+1)}).
\]
By Lemma~\ref{lem:bonuscomparison}, this derivative can switch sign only from positive to
negative. Thus some minimizing cutoff is an endpoint of the union:
either both blocks receive the lower bonus or both receive the higher
bonus. This pooling weakly lowers the bound, preserves monotonicity, and
eliminates one bonus level. Iteration leaves one level throughout
$[a,b]$. In the original cutoff coordinates, every cutoff is then
at $a$ or $b$, a vertex of the ordered-cutoff simplex.

For a general nonnegative monotone schedule, first truncate it as
in Step 1. Approximate the resulting bounded schedule uniformly by
finite-level monotone quantizations. Since $C$ is continuous on
$[0,1]\times[0,\tfrac{1}{2}]$, the corresponding lower bounds of
incentive cost converge. Each quantization admits a constant bonus
with a weakly lower bound. A convergent subsequence of these
constants gives the required constant for the original schedule.
Finally, $\int_a^b C(\type,W_1)\,d\type$ is continuous in $W_1$
on $[0,\tfrac{1}{2}]$, so a minimizing constant exists and weakly
improves on every admissible schedule. Endpoint types have measure
zero, so the same argument applies to open target intervals.

For this target group considered alone, the minimizing constant is
implemented by paying it for state-$1$ evidence at every report
and paying zero otherwise. This contract satisfies \eqref{FD} and
\eqref{IC}, and zero silence payments make \eqref{eq:conditional}
exact on the interval.
The argument for state-$0$ search is symmetric.
\end{proof}

\subsection{A relaxation with a free target cutoff}

\begin{lemma}[A lower bound with a free cutoff]
\label{lem:cutoffrelaxation}
The minimum principal cost over admissible menus is bounded below by
\begin{equation}\label{eq:cutoffrelaxation}
\begin{aligned}
 \min_{0\le\bar\type\le1}\Biggl\{
 &\min_{0\le W_1\le1/2}\int_{\bar\type}^1
   \bigl[\min\{\type(1-q),1-\type\}+\type W_1q\bigr]d\type\\
 {}+{}&\min_{0\le W_1\le1/2}\int_{1-\bar\type}^1
   \bigl[\min\{\type(1-q),1-\type\}+\type W_1q\bigr]d\type
 \Biggr\},\\
 &\text{where}\qquad q=\min\{\type W_1/k,1\}.
\end{aligned}
\end{equation}
Here $\bar\type$ is a cutoff separating search targets, not a
report. Each inner minimization chooses its own constant bonus. The first
integral corresponds to types above the target cutoff $\bar\type$
searching state $1$; the second is the reflected state-$0$ group.
\end{lemma}

\begin{proof}
Lemma~\ref{lem:targetsorting} puts state-$0$ searchers below state-$1$
searchers, with any zero-effort types between them. Choose a cutoff $\bar\type$
in that intervening region and assign its zero-effort types to either
adjacent target with bonus zero. This gives a full partition whose
bonus schedule is nondecreasing in each target's probability.

At each type, \eqref{eq:conditional} is a lower bound on actual cost.
Apply Proposition~\ref{prop:bonuspooling} to each target interval, then allow
the cutoff and the two constant bonuses to be chosen freely. Reflection
of the lower interval uses the uniform density and gives
\eqref{eq:cutoffrelaxation}.

Pure target choices suffice here. At a fixed truthful report, mixing
can occur only between agent best responses. Expected transfers are
linear in the mixing probability, while optimized decision error is
the minimum of two affine functions of that probability. Principal
cost is therefore concave in the mixture, so a pure choice is weakly
better. Selecting such a best response preserves the agent's utility
and reporting constraints. 
The resulting allocation need not itself satisfy reporting incentives:
in particular, its cutoff is independent of its two bonuses. This
only makes its value a lower bound. The proof of
Theorem~\ref{thm:twostructures} establishes feasible attainment
once the cutoff problem below is solved.
\end{proof}

\subsection{The optimal cutoff}

We first establish the curvature property that excludes an
asymmetric interior minimum. Symmetry alone is insufficient.

\begin{lemma}[The capped branch and its curvature]
\label{lem:cappedcurvature}
Suppose $0<k<\tfrac{1}{2}$. On an interval $[\bar\type,1]$ with
$0\le\bar\type<1$, the minimum within the branch with some capped
effort has a unique reward, denoted $\widehat W_1(\bar\type)$, satisfying
\[
 k<\widehat W_1(\bar\type)<\min\{k/\bar\type,1/2\},
\]
where $\tfrac{k}{\bar\type}=+\infty$ at $\bar\type=0$.
The notation $\widehat W_1(\bar\type)$ is used only for the
minimizing constant bonus on this interval; it is not a
report-dependent contract.
It minimizes
\begin{equation}\label{eq:tailcappedcost}
 \frac{W_1}{2}+\frac{k^2(1-W_1)}{6W_1^2}
 -\frac{\bar\type^2}{2}
 +\frac{\bar\type^3W_1(1-W_1)}{3k}.
\end{equation}
For $0<\bar\type<1$, with derivatives taken along this minimizing
reward,
\begin{equation}\label{eq:curvatureimplication}
 \left(\frac{\bar\type^2\widehat W_1(\bar\type)(1-\widehat W_1(\bar\type))}{k}\right)''
 \ge0
 \quad\Longrightarrow\quad
 \left(\frac{\bar\type^2\widehat W_1(\bar\type)(1-\widehat W_1(\bar\type))}{k}\right)'''
 <0.
\end{equation}
\end{lemma}

\begin{proof}
For $k<W_1<\tfrac{k}{\bar\type}$, silence induces action $0$ and effort caps
at $\tfrac{k}{W_1}$. Integrating \eqref{eq:conditional} over
$[\bar\type,\tfrac{k}{W_1}]$ and $[\tfrac{k}{W_1},1]$ gives
\eqref{eq:tailcappedcost}. Its first and second derivatives in $W_1$ are
\begin{align}
 &\frac12+\frac{k^2(W_1-2)}{6W_1^3}
       +\frac{\bar\type^3(1-2W_1)}{3k},
       \label{eq:tailcappedFOC}\\
 &\frac{k^2(3-W_1)}{3W_1^4}-\frac{2\bar\type^3}{3k}
       >\frac{k^2(1-W_1)}{W_1^4}>0.\notag
\end{align}
The first derivative at $W_1=k$ is
$\tfrac{(2k-1)(1-\bar\type^3)}{3k}<0$. At the other endpoint it is
$\tfrac{1-\bar\type^2}{2}>0$ if that endpoint is $\tfrac{k}{\bar\type}$, and
$\tfrac{1}{2}-2k^2>0$ if it is $\tfrac{1}{2}$. This proves existence and uniqueness.
If $W_1\ge \tfrac{k}{\bar\type}$, every type in the interval exerts full effort
and cost is $\tfrac{W_1(1-\bar\type^2)}{2}$, increasing in $W_1$; that branch
cannot improve the minimum.

For the curvature calculation, write $W_1=\widehat W_1(\bar\type)$
and $q=\tfrac{\bar\type W_1}{k}$ for the effort of the
interval's lowest type. Thus $0<W_1<\tfrac{1}{2}$ and $0<q<1$.
Implicit differentiation of \eqref{eq:tailcappedFOC} set equal to zero
gives
\begin{equation}\label{eq:cappedimplicit}
 \bar\type W_1'
 =-\frac{3W_1q^3(1-2W_1)}{3-W_1-2W_1q^3},
 \qquad
 \bar\type(q^3)'
 =\frac{3q^3[3-W_1-(3-4W_1)q^3]}{3-W_1-2W_1q^3}.
\end{equation}
The denominator is positive. To make the sign verification explicit,
repeated differentiation using \eqref{eq:cappedimplicit} yields
\begin{align*}
 &\bar\type\left(\frac{\bar\type^2W_1(1-W_1)}{k}\right)'''
 +\frac{36q^3(1-2W_1)^2}{(3-W_1-2W_1q^3)^2}
       \left(\frac{\bar\type^2W_1(1-W_1)}{k}\right)''\\
 &=-\frac{3W_1q^3(1-2W_1)^2}{k(3-W_1-2W_1q^3)^5}
 \Bigl\{4(3-W_1)^3(W_1+9)(1-q^3)^4\\
 &\hspace{38pt}
 +108(3-W_1)^2(4W_1^2+3W_1+2)q^3(1-q^3)^3\\
 &\hspace{38pt}
 +27[36+176W_1+224W_1^2+W_1^2(1-2W_1)(13+2W_1)]q^6(1-q^3)^2\\
 &\hspace{38pt}
 +54W_1[42+105W_1+76W_1^2+26W_1^2(1-2W_1)]q^9(1-q^3)\\
 &\hspace{38pt}
 +243W_1^2[5+12W_1+4W_1(1-2W_1)]q^{12}\Bigr\}<0.
\end{align*}
Every term inside braces is nonnegative and their sum is positive.
The coefficient of the second derivative on the left is positive.
Consequently, a nonnegative second derivative forces the third
derivative to be strictly negative, proving
\eqref{eq:curvatureimplication}.
\end{proof}

\begin{lemma}[Only a boundary or a symmetric cutoff is needed]
\label{lem:cutoffminimum}
The minimum in \eqref{eq:cutoffrelaxation} is attained at a cutoff
$\bar\type\in\{0,\tfrac{1}{2},1\}$. At $\bar\type=\tfrac{1}{2}$, the two bonuses
can be chosen equal. For $k\ge\tfrac{1}{2}$, a boundary cutoff suffices.
\end{lemma}

\begin{proof}
By reflection restrict attention to $0\le\bar\type\le\tfrac{1}{2}$.
The boundary $\bar\type=0$ is already a single-target allocation.
For an interior cutoff, one interval of target probabilities starts
at $\bar\type$ and the other at $1-\bar\type\ge\tfrac{1}{2}$.

\textbf{Step 1: the majority interval chooses zero or a capped reward.}
Consider the inner minimization on $[\bar\type,1]$ temporarily with
$\tfrac{1}{2}\le\bar\type<1$. For $0<W_1<k$, let
$\type=(1+\sqrt{1-\tfrac{W_1}{k}})^{-1}$ be the type at which the silence action
switches. If this type is below $\bar\type$, differentiating the
lower bound of incentive cost over this interval gives
$\tfrac{2W_1(1-\bar\type^3)}{3k}>0$. If
$\type>\bar\type$, the derivative is
\[
 \frac{2W_1(1-\bar\type^3)+\bar\type^3-\type^3}{3k}.
\]
After division by the positive $W_1$, its numerator has the sign of
\[
 2(1-\bar\type^3)
 -\frac{\type^2(\type^3-\bar\type^3)}{k(2\type-1)},
 \qquad W_1=\frac{k(2\type-1)}{\type^2}.
\]
The fraction being subtracted strictly increases with $\type$.
Indeed, its derivative has the sign of
\begin{align*}
 \type^3(8\type-5)+2\bar\type^3(1-\type)
 &\ge\type^3(8\type-5)+\frac{1-\type}{4}\\
 &=\frac{(2\type-1)^2(8\type^2+3\type+1)}4>0.
\end{align*}
Thus the derivative of cost can cross zero only from positive to
negative on this uncapped branch: there is no interior minimum.

If $k<\tfrac{1}{2}$, the derivative at $W_1=k$ is
$\tfrac{(2k-1)(1-\bar\type^3)}{3k}<0$. The majority interval's optimum
is therefore either $W_1=0$ or the capped minimum in
Lemma~\ref{lem:cappedcurvature}. If $k\ge\tfrac{1}{2}$, the largest allowed
bonus is $\tfrac{1}{2}\le k$. At that endpoint the numerator of the derivative
is $1-\type^3\ge0$ whenever the silence cutoff is in the interval;
otherwise the derivative is positive directly. The preceding
monotonicity then makes cost increasing throughout the allowed
range. Its minimum is $W_1=0$.

\textbf{Step 2: exclude the need for mixed branches.}
Return to $0<\bar\type\le\tfrac{1}{2}$. If the majority interval
$[1-\bar\type,1]$ chooses zero, its cost is $\tfrac{\bar\type^2}{2}$.
For every bonus $W_1\le\tfrac{1}{2}$ and every $\type\le\bar\type$,
\eqref{eq:conditional} is at most $\type$, since
$\type(1-q)+\type W_1q\le\type$. Extending the other interval's
bonus to all types therefore costs no more than the two-interval
allocation. A single target suffices in this case; in particular,
it suffices for every $k\ge\tfrac{1}{2}$.

Otherwise, by Step 1 the majority interval can choose an optimal
capped bonus $W_1'>k$. It weakly beats every uncapped bonus $W_1\le k$
on $[1-\bar\type,1]$. The higher-minus-lower cost difference from
Lemma~\ref{lem:bonuscomparison} crosses only from negative to positive.
If its integral on an upper interval is nonpositive, enlarging that
interval downward preserves a nonpositive integral. Indeed, its
value at $\type=1$ is $W_1'-\tfrac{W_1^2}{k}>0$. A nonpositive integral must
therefore contain a strictly negative difference somewhere in the
original interval. The sign property makes every difference in the
added lower interval nonpositive. Consequently the same $W_1'$ weakly beats each
uncapped $W_1$ on $[\bar\type,1]$ as well. Both intervals can therefore
choose their minimizing capped rewards. Ties do not affect the
argument.

\textbf{Step 3: two capped intervals have no asymmetric minimum.}
For $0<k<\tfrac{1}{2}$, take the sum of the two optimized costs in
\eqref{eq:tailcappedcost}, evaluated at $\bar\type$ and
$1-\bar\type$. Its derivative with respect to $\bar\type$ is
\begin{equation}\label{eq:cappedcutoffslope}
 1-2\bar\type
 +\frac{\bar\type^2\widehat W_1(\bar\type)(1-\widehat W_1(\bar\type))}{k}
 -\frac{(1-\bar\type)^2\widehat W_1(1-\bar\type)(1-\widehat W_1(1-\bar\type))}{k}.
\end{equation}
Here the envelope theorem differentiates only the cutoff in each
optimized cost.

The function $\tfrac{\type^2\widehat W_1(\type)(1-\widehat W_1(\type))}{k}$ has strictly positive
second derivative at zero, namely $\tfrac{2\widehat W_1(0)(1-\widehat W_1(0))}{k}$.
Lemma~\ref{lem:cappedcurvature} implies that its second derivative
strictly decreases wherever it is nonnegative. It can cross zero
at most once, from positive to negative, and cannot return to zero
from below. The continuous endpoint limits of
\eqref{eq:cappedimplicit} give
\[
 \widehat W_1(1)=k,\qquad
 \widehat W_1'(1)=-\frac{k(1-2k)}{1-k},\qquad
 \left.\frac{d}{d\type}
 \frac{\type^2\widehat W_1(\type)(1-\widehat W_1(\type))}{k}\right|_{\type=1}
 =\frac{1-2k^2}{1-k}>1.
\]
The value $\widehat W_1(1)=k$ denotes the limiting minimizing reward as the
interval shrinks to zero length.

If the second derivative of $\tfrac{\type^2\widehat W_1(\type)(1-\widehat W_1(\type))}{k}$ at
$\type=\bar\type<\tfrac{1}{2}$ is nonnegative, it is strictly greater than
the second derivative at $\type=1-\bar\type$. This follows either
from its strict decrease while nonnegative or from the latter
second derivative already being negative. Thus the derivative in
\eqref{eq:cappedcutoffslope} is strictly convex in this region.

If instead that second derivative is negative at $\bar\type$,
it stays negative on $[\bar\type,1]$. The first derivative of
$\tfrac{\type^2\widehat W_1(\type)(1-\widehat W_1(\type))}{k}$ then decreases toward its endpoint
value, which is greater than one. Hence
\[
 1-2\bar\type
 -\int_{\bar\type}^{1-\bar\type}
     \frac{d}{d\type}
       \frac{\type^2\widehat W_1(\type)(1-\widehat W_1(\type))}{k}\,d\type<0.
\]
This is exactly \eqref{eq:cappedcutoffslope}.

The slope in \eqref{eq:cappedcutoffslope} starts at $k>0$ when
$\bar\type=0$ and equals zero at $\bar\type=\tfrac{1}{2}$. It is strictly
convex until any change of curvature described above, and negative
afterward. It can therefore change sign in $(0,\tfrac{1}{2})$ only from
positive to negative. The sum of optimized capped costs has no
asymmetric interior minimum and is minimized at $0$ or $\tfrac{1}{2}$.

Finally, Step 2 shows that every cutoff either is weakly dominated
by a single target or has value equal to the sum of capped minima.
The latter sum is bounded below by its endpoint values just found.
These observations also cover changes between the uncapped, capped,
and zero-reward branches. At the symmetric cutoff the two inner
problems are identical, so their minimizing bonuses can be chosen
equal. Reflecting the argument adds the cutoff $1$.
\end{proof}

\section{Proof of Theorem \ref{thm:twostructures}}
\label{app:theorem}

\subsection{Reduction and attainment}

Lemma~\ref{lem:targetsorting} orders the target groups and their
bonuses. Applying Proposition~\ref{prop:bonuspooling} within each
group gives the free-cutoff lower bound of incentive cost in
Lemma~\ref{lem:cutoffrelaxation}. Lemma~\ref{lem:cutoffminimum}
shows that its minimum occurs at a boundary cutoff or at
$\tfrac{1}{2}$.

These minimizing allocations are implementable. At cutoff zero,
pay only for state-$1$ evidence; every type searches state $1$ and
chooses $q(\type)=\min\{\tfrac{\type W_1}{k},1\}$. Cutoff one is
the reflected case. At cutoff $\tfrac{1}{2}$, equal bonuses induce
search for the more likely state, with effort
$q(\type)=\min\{\tfrac{W_1\max\{\type,1-\type\}}{k},1\}$.
Because payments do not depend on the report, truthful reporting
satisfies \eqref{IC}. The principal matches every disclosed
certificate and chooses optimally after silence. Zero silence
payments make \eqref{eq:conditional} and its state-$0$ counterpart
equal actual cost at every type. Optimizing within the two payment
structures and choosing the cheaper one therefore attains the
minimum lower bound of incentive cost.

The optimal contract induces positive acquisition. Any contract
with no acquisition has expected decision loss
$\int_0^1\min\{\type,1-\type\}\,d\type=\tfrac14$ and nonnegative
payments. In contrast, paying a constant
$0<W_1<\min\{k,\tfrac18\}$ for state-$1$ evidence and zero otherwise
satisfies \eqref{IC} and induces $q(\type)=\tfrac{\type W_1}{k}$.
Evaluating this contract with the prior-majority action after
silence gives the upper bound
\[
 \frac14+\frac{W_1(8W_1-1)}{24k}<\frac14.
\]
The principal's optimal silence action can only lower this cost.
Thus zero acquisition is strictly suboptimal. It remains to
optimize the positive bonuses and compare the two structures.
Throughout this proof, cost means expected decision loss plus
expected wage payments, the negative of the objective in \eqref{P}.

\subsection{Optimal bonus for one certificate}

Take $W_0=0$ and $W_1>0$. By \eqref{IC}, every type searches state $1$
and chooses
\begin{equation*}
 q(\type)=\min\left\{\frac{\type W_1}{k},1\right\}.
\end{equation*}
Conditional on the prior $\type$, the joint probabilities of silence
and states $1$ and $0$ are $\type(1-q(\type))$ and $1-\type$.
The principal chooses the action that makes the smaller of these
probabilities erroneous. His cost is therefore
\begin{equation}\label{eq:single_cost_integral}
 \int_0^1\left[
 \min\{\type(1-q(\type)),1-\type\}
 +\type W_1q(\type)\right]d\type,
 \qquad 0<W_1\le\frac12.
\end{equation}

\paragraph{Interior effort.}
If $W_1<k$, all types choose $q(\type)=\tfrac{\type W_1}{k}<1$.
The two silence actions have equal loss when
$\type(1-\tfrac{\type W_1}{k})=1-\type$. Hence the optimal silence action is
\begin{equation*}
 a(\hat\type,\emptyset)=
 \begin{cases}
 0,&\displaystyle
 \hat\type\le\frac{1}{1+\sqrt{1-W_1/k}},\\[4pt]
 1,&\displaystyle
 \hat\type>\frac{1}{1+\sqrt{1-W_1/k}}.
 \end{cases}
\end{equation*}
The principal always matches a disclosed certificate. Integrating
\eqref{eq:single_cost_integral} gives the interior-effort cost
\begin{equation}\label{eq:interiorcost}
 \frac12-\frac{2}{3(1+\sqrt{1-W_1/k})}
 +\frac{1}{3(1+\sqrt{1-W_1/k})^2}
 +\frac{W_1^2}{3k}.
\end{equation}
The first three terms are decision loss; the last is the expected
wage. Differentiation gives
\[
 \frac{2W_1-(1+\sqrt{1-W_1/k})^{-3}}{3k}.
\]
The moving action cutoff contributes no additional term because
the two actions have equal loss there. An interior minimum therefore
satisfies \eqref{eq:thirdinterior}. To classify the stationary
points, write the silence cutoff as $\type$. At a stationary point,
\eqref{eq:thirdinterior} gives
\[
 k=\frac{\type^5}{2(2\type-1)},
\]
The right-hand side decreases on $(\tfrac12,\tfrac58)$ and
increases on $(\tfrac58,1)$, attaining its minimum
$\tfrac{3125}{16384}$ at $\type=\tfrac58$. Below this value of $k$,
the uncapped objective is strictly decreasing; at equality, its
stationary point has no sign change. For
$\tfrac{3125}{16384}<k<\tfrac12$, the smaller stationary bonus is
the sole interior minimum and the larger is a maximum. The
minimum's silence cutoff lies below $\tfrac58$. For $k\ge\tfrac12$,
only the smaller stationary root is relevant on the admissible
bonus range. Whenever the capped minimum also exists, it must be
compared with this interior minimum.

\paragraph{Some effort is capped.}
If $k<W_1<\tfrac{1}{2}$, types below $\tfrac{k}{W_1}$ exert interior effort and types
above it choose $q=1$. Silence induces action $0$ for every report.
For an uncapped type,
$2\type-\tfrac{\type^2W_1}{k}\le \tfrac{k}{W_1}<1$, so the loss from action $0$ is
smaller; for a capped type, silence rules out state $1$.
The decision loss and expected wage are
\[
 \Pr(a\ne\state)=\frac{k^2}{6W_1^2},
 \qquad
 \E[w(\type,\cert)]=\frac{W_1}{2}-\frac{k^2}{6W_1}.
\]
Thus the cost on this branch is
\begin{equation}\label{eq:cappedcost}
 \frac{W_1}{2}+\frac{k^2(1-W_1)}{6W_1^2}.
\end{equation}
Its first-order condition is \eqref{eq:thirdcapped}.
The second derivative of cost is
$\tfrac{k^2(3-W_1)}{3W_1^4}>0$. For $0<k<\tfrac{1}{2}$, the first derivative is
negative at $W_1=k$ and positive at $W_1=\tfrac{1}{2}$, so
\eqref{eq:thirdcapped} has a unique minimizing root in $(k,\tfrac{1}{2})$.
The costs \eqref{eq:interiorcost} and \eqref{eq:cappedcost} agree
at $W_1=k$. For $k\ge\tfrac{1}{2}$, the derivative of
\eqref{eq:interiorcost} is positive just below $W_1=\tfrac{1}{2}$, including
when $k=\tfrac{1}{2}$, so this upper endpoint is also suboptimal. Consequently, when both stationary minima exist, the best
one-certificate contract is found by comparing those two minima;
the common branch boundary is strictly dominated by the capped
minimum. When only one minimum exists, it is optimal within this
payment structure. Every optimal positive bonus is strictly below
$\tfrac{1}{2}$.

\subsection{Optimal equal rewards}

Now let $W_0=W_1>0$. The agent searches her more likely state,
because it generates the reward more often. Condition~\eqref{IC}
gives
\begin{equation*}
 q(\type)=\min\left\{
 \frac{W_1\max\{\type,1-\type\}}{k},1\right\}.
\end{equation*}
For $\type\ge\tfrac{1}{2}$, select target $1$; the lower half of types is
symmetric. Expected cost is
\begin{equation}\label{eq:symmetric_cost_integral}
 2\int_{1/2}^1\left[
 \min\{\type(1-q(\type)),1-\type\}
 +\type W_1q(\type)\right]d\type.
\end{equation}

\paragraph{The relevant effort range.}
For $0<W_1<k$, the derivative of \eqref{eq:symmetric_cost_integral}
has the sign of
\[
 14k-\frac{4}{(1+\sqrt{1-W_1/k})^4}
       -\frac{2}{(1+\sqrt{1-W_1/k})^3}
       -\frac{1}{(1+\sqrt{1-W_1/k})^2}.
\]
This expression strictly decreases with $W_1$, so a stationary
point in this range can only be a maximum. For $k\ge\tfrac{1}{2}$, cost is
increasing throughout the admissible range $0<W_1\le\tfrac{1}{2}$, and zero
bonuses are preferable within this structure.

If $W_1\ge2k$, all types exert full effort, decision loss is zero,
and cost is $\tfrac{3W_1}{4}$, increasing in $W_1$. For $0<k<\tfrac{1}{2}$, a
positive optimum must therefore lie in
$k<W_1<\min\{2k,\tfrac{1}{2}\}$. On this branch, effort is capped for the
types whose more likely state has probability at least $\tfrac{k}{W_1}$.
Silence induces the opposite of the searched state: action $1$ below
$\tfrac{1}{2}$ and action $0$ at or above $\tfrac{1}{2}$.

\paragraph{The optimal positive bonus.}
Integrating over the uncapped and capped types gives
\begin{align*}
 \Pr(a\ne\state)&=\frac{k^2}{3W_1^2}-\frac14+\frac{W_1}{12k},\\
 \E[w(\type,\cert)]&=W_1-\frac{k^2}{3W_1}
                         -\frac{W_1^2}{12k}.
\end{align*}
Their sum is
\begin{equation}\label{eq:symmetriccost}
 W_1-\frac14+\frac{W_1(1-W_1)}{12k}
             +\frac{k^2(1-W_1)}{3W_1^2}.
\end{equation}
The first-order condition reduces to \eqref{eq:thirdsymmetric}.
For $0<k<\tfrac{1}{2}$, it has a unique minimizing root in
$(k,\min\{2k,\tfrac{1}{2}\})$. Indeed, the second derivative
of \eqref{eq:symmetriccost} is
\[
 -\frac{1}{6k}+\frac{2k^2(3-W_1)}{3W_1^4}>0
\]
on this interval, and the first derivative is negative at its lower
endpoint and positive at its upper endpoint. The resulting cost must
be compared with zero bonuses and the contract rewarding one certificate.
A positive optimum within this structure is also strictly below $\tfrac{1}{2}$.

\subsection{The two parameter thresholds}

For this comparison only, write $S(k)$, $C(k)$, and $I(k)$ for
\eqref{eq:symmetriccost}, \eqref{eq:cappedcost}, and
\eqref{eq:interiorcost}, respectively, evaluated at their minimizing
stationary bonuses. The positive equal-reward candidate $S(k)$ and
the capped one-certificate candidate $C(k)$ exist for
$0<k<\tfrac{1}{2}$. The interior one-certificate minimum $I(k)$
exists for $k>\tfrac{3125}{16384}$. The preceding calculations
exclude all other positive candidates. Zero bonuses are strictly
dominated by the small one-certificate reward constructed above.

The thresholds are defined by
\begin{equation}\label{eq:thresholds}
 S(k_1)=C(k_1),\qquad C(k_2)=I(k_2).
\end{equation}
We establish that each equality has exactly one solution in the
relevant range and that $k_1<k_2$.

\paragraph{The first threshold: equal rewards versus one capped reward.}
Within this paragraph, let $r$ and $t$ be the ratios of $k$ to the
minimizing bonuses in $C(k)$ and $S(k)$, respectively. Their
first-order conditions give
\[
 k=\frac{2r^3}{3+r^2},\qquad 0<r<1,
 \qquad
 k=\frac{t(8t^3-1)}{2(2t^3+6t-1)},\qquad \frac12<t<1.
\]
Both right-hand sides are strictly increasing in their respective
arguments: their derivatives are
\[
 \frac{2r^2(9+r^2)}{(3+r^2)^2}>0,
 \qquad
 \frac{16t^6+4t^3(36t-7)+1}{2(2t^3+6t-1)^2}>0.
\]
Expressing the two objectives in these ratios gives
\[
 \begin{aligned}
 C(k)&=\frac{r^2}{6}+k\frac{3-r^2}{6r},\\
 S(k)&=\frac{(2t-1)^2(t+1)}{12t}
       +k\left(\frac1t-\frac1{12t^2}-\frac t3\right).
 \end{aligned}
\]
The envelope theorem therefore implies
\[
 \frac{d}{dk}\left(\frac{S(k)-C(k)}{k}\right)
 =\frac1{k^2}\left[
 \frac{r^2}{6}-\frac{(2t-1)^2(t+1)}{12t}\right].
\]
The expression in brackets is strictly positive. To verify this,
set $z=\tfrac{(2t-1)(t+1)}{2t}\in(0,1)$ and write
\[
 P(t)=8t^6+12t^5+t^2(66t^2-9)+t(49t^2-12)+2>0.
\]
Every term is positive for $\tfrac{1}{2}<t<1$. Direct calculation gives
\[
 \frac{t(8t^3-1)}{2(2t^3+6t-1)}-\frac{2z^3}{3+z^2}
 =\frac{(1-t)(2t-1)P(t)}
 {2t(2t^3+6t-1)(4t^4+4t^3+9t^2-2t+1)}>0.
\]
Since $\tfrac{2r^3}{3+r^2}$ increases with $r$,
this inequality implies $r>z$. Also,
\[
 z^2\ge\frac{(2t-1)^2(t+1)}{2t}.
\]
Thus $\tfrac{S(k)-C(k)}{k}$ strictly increases throughout
$(0,\tfrac{1}{2})$.

As $k\downarrow0$, $\tfrac{S(k)}{k}\to\tfrac32$ whereas
$\tfrac{C(k)}{k}\to\infty$. As $k\uparrow\tfrac{1}{2}$,
$S(k)\to\tfrac{11}{24}$ and $C(k)\to\tfrac13$.
Consequently there is exactly one crossing $k_1$, with $S(k)<C(k)$
below it and $S(k)>C(k)$ above it.

An exact upper bound places this crossing before an interior
one-certificate minimum can arise. At $t=\tfrac23$, the
equal-reward first-order condition gives $k=\tfrac{37}{291}$ and
$S(k)=\tfrac{6779}{41904}$. Evaluating the one-certificate objective
at the feasible ratio $r=\tfrac35$ yields
$C(k)\le\tfrac{6689}{43650}$, so
\[
 S(k)-C(k)\ge\frac{8939}{1047600}>0,
 \qquad
 0<k_1<\frac{37}{291}<\frac{3125}{16384}.
\]
For $k<k_1$, $C(k)$ is therefore the best one-certificate cost, and
equal rewards are optimal. For $k_1<k<\tfrac{1}{2}$, the positive
equal-reward candidate is worse than $C(k)$ and hence worse than
the best one-certificate contract. Zero rewards are also dominated.
For $k\ge\tfrac{1}{2}$, zero rewards minimize cost within the
equal-reward family and are again dominated by a one-certificate
contract. Equal rewards thus never become optimal again.

\paragraph{The second threshold: capped versus interior effort.}
For this paragraph, let $r$ again be the ratio of $k$ to the capped
one-certificate bonus, and let $t$ be the silence-action cutoff at
the interior minimum. Then
\[
 k=\frac{2r^3}{3+r^2},\qquad 0<r<1,
 \qquad
 k=\frac{t^5}{2(2t-1)},\qquad \frac12<t<\frac58.
\]
The minimized costs and their envelope derivatives are
\[
 \begin{aligned}
 C(k)&=\frac{r^2(9-r^2)}{6(3+r^2)},
 &C'(k)&=\frac{3-r^2}{6r}>\frac13,\\
 I(k)&=\frac{3-5t+4t^2}{6},
 &I'(k)&=\frac{(2t-1)^2}{3t^4}<\frac13.
 \end{aligned}
\]
Thus $C(k)-I(k)$ is strictly increasing wherever both candidates
exist. At $k=\tfrac14$, the feasible capped bonus $W_1=\tfrac13$
gives
\[
 C\left(\frac14\right)\le\frac{11}{48}
 <\frac{23}{96}\le I\left(\frac14\right).
\]
At $k=\tfrac13$, the left-hand side of \eqref{eq:thirdcapped},
evaluated at $W_1=\tfrac38$, is $-\tfrac{103}{4608}<0$.
The minimizing capped bonus therefore exceeds $\tfrac38$.
Using its first-order condition, its cost can be written as
$\tfrac{W_1(3-2W_1)}{2(2-W_1)}$, which is strictly increasing in
$W_1\in(0,\tfrac12)$. Hence
\[
 C\left(\frac13\right)>\frac{27}{104}
 >\frac14>I\left(\frac13\right).
\]
There is exactly one crossing
$k_2\in(\tfrac14,\tfrac13)$, and in particular $k_1<k_2$.
Within the one-certificate family, the capped bonus is optimal below
$k_2$; the interior bonus is optimal above it. For $k\ge\tfrac12$, only the
interior minimum remains relevant. At $k=k_2$, both bonuses attain
the same minimum.

Solving the two unique equalities in \eqref{eq:thresholds}
numerically gives the values reported in the theorem. The regime
ordering follows from the preceding inequalities, independently of
this numerical evaluation. Together with the reduction and
attainment argument, this completes the proof of
Theorem~\ref{thm:twostructures}.\hfill\qedsymbol

\end{document}